\documentclass[article,aps,amsmath,amsfonts,superscriptaddress,a4paper,10pt,twocolumn,showpacs]{revtex4}

\usepackage{graphics}
\newcommand \beq{\begin{equation}}
\newcommand \beqa{\begin{eqnarray}}
\newcommand \beqann{\begin{eqnarray*}}
\newcommand \eeq{\end{equation}}
\newcommand \eeqa{\end{eqnarray}}
\newcommand \eeqann{\end{eqnarray*}}

\newcommand{\Jxe}{J_x^{\textrm{eff}}}
\newcommand{\Jye}{J_y^{\textrm{eff}}}

\newcommand{\beal}{\begin{align}}
\newcommand{\enal}{\end{align}}
\newcommand{\beeq}{\begin{equation}}
\newcommand{\eneq}{\end{equation}}

\usepackage{ulem}
\usepackage[usenames]{color}

\newcommand\rsout{\bgroup\markoverwith
{\textcolor{red}{\rule[0.5ex]{2pt}{0.4pt}}}\ULon}

\begin{document}
\title{Floquet analysis of the modulated two-mode Bose-Hubbard model}
\author{Gentaro Watanabe}
\affiliation{Asia Pacific Center for Theoretical Physics (APCTP), San 31, Hyoja-dong, Nam-gu, Pohang, Gyeongbuk 790-784, Korea}
\affiliation{Department of Physics, POSTECH, San 31, Hyoja-dong, Nam-gu, Pohang, Gyeongbuk 790-784, Korea}
\affiliation{Nishina Center, RIKEN, 2-1 Hirosawa, Wako, Saitama 351-0198, Japan}

\author{Harri M\"akel\"a}
\affiliation{Department of Physics, Ume\aa \,\,University, SE-901 87 Ume\aa, Sweden}
\affiliation{Department of Applied Physics/COMP, Aalto University, P.O. Box 14100, FI-00076 AALTO, Finland}

\begin{abstract}
We study the tunneling dynamics in a time-periodically modulated
two-mode Bose-Hubbard model using Floquet theory. We consider
situations where the system is in the self-trapping regime and either
the tunneling amplitude, the interaction strength, or the energy
difference between the modes is modulated. In the former two cases,
the tunneling is enhanced in a wide range of modulation
frequencies, while in the latter case the resonance is narrow.  We
explain this difference with the help of Floquet analysis.  If the
modulation amplitude is weak, the locations of the resonances can be
found using the spectrum of the non-modulated Hamiltonian.
Furthermore, we use Floquet analysis to explain the coherent
destruction of tunneling (CDT) occurring in a large-amplitude
modulated system.  Finally, we present two ways to create a NOON
state (a superposition of 
$N$ particles in mode 1 with zero particles in mode 2 and vice versa).
One is based on a coherent oscillation caused by detuning from a partial CDT. 
The other makes use of an adiabatic variation of the
modulation frequency.  This results in a Landau-Zener type of
transition between the ground state and a NOON-like state.
\end{abstract}


\maketitle

\section{Introduction}

Ultracold atomic gases are novel systems with a high degree of
controllability, making them very useful in the studies on quantum
phenomena.
The possibility to control the
parameters during experiments is essential, for example, in quantum
information processing (e.g., Refs. \cite{Hammerer,Saffman}) and
matter-wave interferometry (e.g.,
Refs.\ \cite{cronin09,bongs04,lee12}).  In this paper, we discuss the
dynamics of an ultracold bosonic gas trapped in a time-periodically
modulated potential. The dynamics of periodically modulated quantum
systems has attracted both theoretical (e.g.,
Refs.\ \cite{dunlap,cdt,holthaus,grifoni,abdullaev00,holthaus01,haroutyunyan01,lee01,salmond,kohler03,haroutyunyan,eckardt,wang05,creffield07,creffield08,weiss,strzys08,luo08,witthaut09,wang09,gong09,xie09,catform,kolovsky,kudo,swallowtail,brouzos11})
and experimental (e.g.,
Refs.\ \cite{gemelke,lignier,gommers08,sias,alberti,eckardt09,zenesini09,creffield10,alberti2})
interest during recent years.  It is known that a modulated system has
resonances at which the tunneling is either suppressed or enhanced. In
the neighborhood of a resonance, the behavior of the system is very
sensitive to the modulation frequency.

The suppression of tunneling by modulating the energy difference
between the modes is known as the coherent destruction of tunneling (CDT) \cite{dunlap,cdt,holthaus}.  
CDT was discovered in Ref.\ \cite{dunlap}, where the
motion of a charged particle in a lattice under the influence of an oscillating
electric field was studied. It was shown that an
initially localized particle remains localized in a one-dimensional
lattice if the amplitude and frequency of the electric field are
chosen suitably. In Ref.\ \cite{cdt}, CDT was found to
occur in systems consisting of a particle subjected to a
periodic force and trapped in a double-well potential. Recently, the
coherent destruction of tunneling has been actively studied in the
context of ultracold bosonic atoms (e.g., Refs.\ \cite{haroutyunyan,eckardt,creffield08,strzys08,luo08,gong09,lignier,sias,eckardt09,zenesini09,creffield10}).

Unlike the CDT, which is typically observed under the condition that
the tunneling coupling is larger than or comparable to the interaction
energy, the enhancement of tunneling by modulation can take place in a
system where the interaction energy dominates over the tunneling
coupling.  In the absence of modulation, the large interaction energy
suppresses tunneling for energetic reasons
\cite{milburn97,smerzi97}. This leads to a very long tunneling period
(the time needed for $N$ particles to tunnel from one mode to another
and back).  However, by modulating the tunneling matrix element, it is
possible to enhance the many-particle tunneling and thereby reduce the
tunneling period \cite{catform}. In this paper, we analyze the reasons
behind the enhanced tunneling with the help of a detailed Floquet
analysis.  In order to make the analysis more complete, we consider
also systems where either the interaction strength or the energy
difference between the modes is modulated.  We find that these two
methods provide an alternative way to enhance tunneling.  It is shown
that the width of the resonance, that is, the range of modulation
frequencies corresponding to the enhanced tunneling, depends strongly
on whether the tunneling matrix element, the interaction strength, or
the energy difference between the two modes is modulated; the
resonance is wider in the former two cases. We explain this difference
with the help of Floquet theory and the eigenvalues of the
non-modulated Hamiltonian.

We analyze also the coherent destruction of tunneling using Floquet
theory. It is known that CDT can be caused by modulating either the
energy difference between the modes or the interaction strength.  We
study only the Floquet spectrum of the former system because it has
not received much attention in the literature, unlike the Floquet
spectrum of the interaction-modulated system \cite{strzys08,gong09}. 
In addition to this, we present two ways to
generate NOON-like (Schr\"odinger's-cat--like) states.  The first is
based on the CDT induced by a large-amplitude modulation of
interaction strength, whereas the second makes use of a
small-amplitude modulation of the tunneling coupling.

This paper is organized as follows. In Sec.\ \ref{sec_BH}, we
define the modulated Bose Hubbard Hamiltonian. In
Sec.\ \ref{sec_Floquet}, we give a short summary of
the Floquet theory used in this article. Section \ref{sec_selftrapping}
discusses in depth the results of the Floquet analysis for systems
in the self-trapping regime subjected to a small-amplitude modulation.  
In Sec.\ \ref{sec_Josephson}, the
coherent destruction of tunneling is examined using Floquet theory.  
It is also shown that NOON states can be created with the help of partial CDT.
In Sec.\ \ref{sec_noon}, a way to create NOON states using adiabatic
sweep across an avoided crossing is presented.  Finally, the
conclusions are in Sec.\ \ref{sec_conclusions}.

\section{Time-periodically modulated two-site Bose Hubbard Hamiltonian}\label{sec_BH}

We consider a system described by the two-mode Bose-Hubbard
Hamiltonian.  For definiteness, we assume that this model is realized
physically by bosons trapped in a double-well potential.  We consider
situations where either the tunneling amplitude, the interaction
strength, or the energy difference between the wells is modulated
periodically in time. This system is described by the Hamiltonian
\begin{eqnarray}
\nonumber 
\hat{H}(t) &=& -J(t) (\hat{c}_1^\dagger \hat{c}_2 + \hat{c}_2^\dagger \hat{c}_1)
+ \frac{U(t)}{2} (\hat{c}_1^\dagger\hat{c}_1^\dagger\hat{c}_1\hat{c}_1 
+ \hat{c}_2^\dagger\hat{c}_2^\dagger\hat{c}_2\hat{c}_2)\\
&&+ \frac{V(t)}{2} (\hat{c}_1^\dagger\hat{c}_1 - \hat{c}_2^\dagger\hat{c}_2)\\
\label{eq_H}
&=& -2J(t) \hat{S}_x + U(t) \hat{S}_z^2 + V(t) \hat{S}_z .
\label{eq_h}
\end{eqnarray}
Here $J$ is the tunneling matrix element, $U$ is the on-site interaction, and
$V$ is the energy difference between the wells (tilt).
We have introduced the SU(2) generators defined as
\begin{eqnarray}
  \hat{S}_x &=& \frac{1}{2} (\hat{c}_1^\dag\hat{c}_2+\hat{c}_2^\dag\hat{c}_1)\ ,\\
  \hat{S}_y &=& \frac{1}{2i} (\hat{c}_1^\dag\hat{c}_2-\hat{c}_2^\dag\hat{c}_1)\ ,\\
  \hat{S}_z &=& \frac{1}{2} (\hat{c}_1^\dag\hat{c}_1-\hat{c}_2^\dag\hat{c}_2)\ ,
\end{eqnarray}
where $\hat{c}_i (\hat{c}_i^\dag)$ annihilates (creates) an atom in mode $i$.

We define the time-dependent tunneling matrix  element as 
\begin{equation}
  J(t) = J_0 [1+A_J \sin{(\omega t + \phi_J)}],
\label{eq_jt}
\end{equation}
where $J_0$ is the amplitude of the time-independent part and 
$A_J \in [0,1]$ gives the relative amplitude of the 
time-dependent tunneling matrix element. The modulated tilt and interaction strength 
are defined as 
\begin{align}
  U(t) &= U_0+ U_1 \sin{(\omega t + \phi_U)},\label{eq_ut}\\  
  V(t) &= V_0+ V_1 \sin{(\omega t + \phi_V)},
\label{eq_vt}
\end{align}
where $U_0,V_0$ ($U_1,V_1$) are the amplitudes of the static (time-dependent) part of the interaction and the tilt, respectively. In the above equations, $\omega$ is the modulation frequency and 
$\phi_J,\phi_U$, and $\phi_V$ are the phase offsets.  
In this paper, time is measured in units of 
\begin{align}
T_0=\frac{\pi}{J_0},
\end{align}
which is the tunneling period in the absence of the interaction ($U_0=U_1=0$) and the tilt ($V_0=V_1=0$).
Here and in what follows, we set $\hbar=1$.

\section{Floquet operator}\label{sec_Floquet}

If the Hamiltonian $\hat{H}$ is periodic in time, 
Floquet theory provides a powerful tool to analyze the dynamics of the system. 
In the following, we denote the modulation period of $\hat{H}$ by $T_\omega$. 
In our case, the modulation is sinusoidal and hence $T_\omega=2\pi/\omega$. 
According to the Floquet theorem (see, e.g., Ref.\ \cite{Chicone}), the time-evolution 
operator $\hat{U}_{\hat{H}}$ determined by the Hamiltonian of Eq. (\ref{eq_H}) 
can be written as 
\begin{align}
\hat{U}_{\hat{H}}(t) =\hat{M}(t) e^{-i t \hat{K}},
\end{align}
where $\hat{M}$ is a periodic matrix with minimum period $T_\omega$ 
and $\hat{M}(0)=\hat{\textrm{I}}$ and $\hat{K}$ is a time-independent operator. 
We define the Floquet operator  $\hat{F}$ as 
\begin{align}
\hat{F} &=\hat{U}_{\hat{H}}(T_\omega)\\
&=\mathcal{T}\left\{\exp{\left[-i\int_0^{T_\omega} \hat{H}(t) dt\right]}\right\}, 
\end{align} 
where $\mathcal{T}$ is the time-ordering operator. 
At times $t=nT_\omega$, where $n$ is an integer, 
we get $\hat{U}_{\hat{H}}(nT_\omega)=e^{-i n T_\omega \hat{K}}=\hat{F}^n$. 
The Floquet operator is a mapping between the state 
at $t=0$ and the state after one modulation period $T_\omega=2\pi/\omega$:
$\Psi(T_\omega) = \hat{F} \Psi(0)$.
The columns of the Floquet operator $\hat{F}$ can be obtained
by following the time evolution of the basis states for one modulation period.
Each time-evolved basis state forms a column of the matrix $\hat{F}$. 
The Hilbert space of a two-mode system containing $N$ bosons is $\mathbb{C}^{N+1}$.  The basis of this Hilbert space can be chosen to be   
$\{|\Delta N\rangle\, ;\, \Delta N =-N,-N+2,-N+4,\ldots, N\}$, where  
 $|\Delta N\rangle$ is a state with $(N+\Delta N)/2$ particles in mode $1$ 
 and $(N-\Delta N)/2$ particles in mode $2$. Any pure state of the system can be written as 
\begin{equation}
\psi =\sum_{\Delta N=-N}^N\, C_{\Delta N} |\Delta N\rangle,
\end{equation} 
where the amplitudes $\{ C_{\Delta N}\}$ are
complex numbers. If $N$ is even (odd), $\Delta N$ takes only even (odd) values.

In order to characterize the eigenstates of the Floquet operator, 
we define the parity operator $\hat{P}$ as 
\begin{align}
\hat{P}|\Delta N\rangle =|-\Delta N\rangle.
\end{align}
It can alternatively be written as $\hat{P}= (-i)^N e^{i\pi \hat{S}_x}$.
 The eigenvalues of $\hat{P}$ are $1$ and $-1$, corresponding to even and odd parity, respectively.  
Because $\hat{P}^\dagger \hat{S}_z\hat{P}=-\hat{S}_z$, the Hamiltonian, and consequently the time-evolution 
operator, commutes with $\hat{P}$  
if the tilt vanishes. Then the eigenstates of $\hat{F}$ are also eigenstates of $\hat{P}$ and  
either $C_{\Delta N} =C_{-\Delta N}$ or $C_{\Delta N} =-C_{-\Delta N}$  
holds for the components of the eigenvectors of $\hat{F}$. 
In the former case, the eigenstate has even parity and is said to be  
symmetric, while in the latter case the parity is odd and 
the eigenstate is called antisymmetric. Furthermore, the absolute values of the 
coefficients $\{C_{\Delta N}\}$ of an eigenstate have maxima at $\Delta N=\pm k$, where $k\geq 0$ is an integer. We denote such an eigenstate by $\psi_k^{(\pm)}$, where $+$ $(-)$ indicates that the eigenvector is symmetric (antisymmetric).  
If $J_0\ll U_0N$ and the signs of the eigenvectors are defined appropriately, we get 
\begin{align}
\psi_N^{(\pm)}\approx \frac{1}{\sqrt{2}}
\left(|N\rangle \pm |-N\rangle\right),
\end{align}
which is valid to zeroth order in $J_0/U_0N$.  For non-zero $V_0$ or
$V_1$, the Floquet eigenstates are neither exactly symmetric nor
antisymmetric because $\hat{S}_z$ is not invariant under the parity
operator.  However, since the time average of the $\hat{S}_z$ term is
zero (we assume that $V_0=0$), the Floquet eigenstates are almost
symmetric or antisymmetric, provided $V_1$ is small. We thus use the
notation $\psi_i^{(\pm)}$ also in this case. Note that, in the case of
large-amplitude modulation of the tilt, the Floquet eigenstates cannot
be classified in this way.

The Floquet operator is a unitary 
operator, and therefore the eigenvalue corresponding to the eigenvector 
 $\psi_{i}^{(\pm)}$ can be written as $e^{i \phi_{i}^{(\pm)}}$. 
The eigenvalue equation becomes  
\begin{equation}
\hat{F}\psi_{i}^{(\pm)} =e^{i \phi_{i}^{(\pm)}}\psi_{i}^{(\pm)}.
\end{equation}
In this paper, we call $\phi_{i}^{(\pm)}\in [-\pi,\pi)$ the phase of a Floquet eigenvalue.

Assume that the initial state is $|N\rangle\approx (\psi_N^{(+)}+\psi_N^{(-)})/\sqrt{2}$. 
At $t=nT_\omega$ ($n\in \mathbb{N}$), the state reads 
\begin{align}
\hat{F}^n|N\rangle \approx\frac{e^{i n\phi_N^{(+)}}}{\sqrt{2}} \left(
 \psi_N^{(+)}+ e^{i n[\phi_N^{(-)}-\phi_N^{(+)}]}\psi_N^{(-)}\right).
\end{align}
If $ n|\phi_N^{(-)}-\phi_N^{(+)}|\approx\pi$, we get
$\hat{F}^n|N\rangle\approx |-N\rangle$; that is, the system has
tunneled from $|N\rangle$ to $|-N\rangle$.  In this paper, we define
the tunneling period as the time needed for the system to tunnel from
$|N\rangle$ to $|-N\rangle$ and back. In terms of the phases of the
Floquet eigenvalues, the tunneling period reads
\begin{align}
T\approx \frac{2\pi T_\omega}{ |\phi_N^{(-)}-\phi_N^{(+)}|}.
\end{align}
Increasing $|\phi_N^{(-)}-\phi_N^{(+)}|$ reduces the tunneling period and vice versa. When $\phi_N^{(+)}=\phi_N^{(-)}$, the tunneling period diverges.

\section{Tunneling period and Floquet analysis in the self-trapping regime}\label{sec_selftrapping}

\begin{figure}
\begin{center}
\rotatebox{0}{
\resizebox{8.2cm}{!}
{\includegraphics{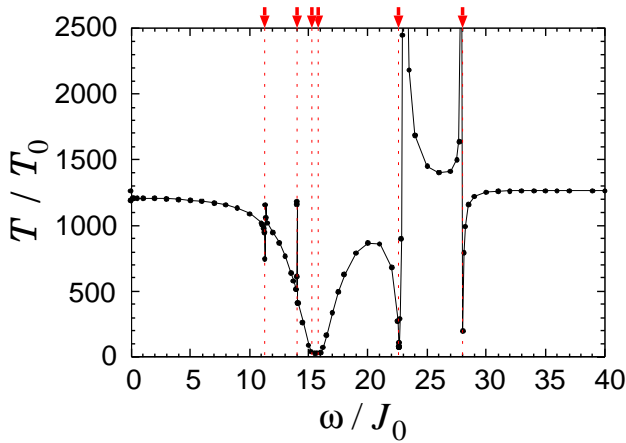}}}
\caption{\label{fig_modj}(Color online)
Tunneling period $T$ in the case of modulated tunneling matrix element $J$ 
for $N = 5$, $U_0/J_0 = 4$, and $A_J = 0.1$ 
(and $V_0=V_1=U_1=0$).
We have set $\phi_J=0$, but there are 
no noticeable differences for different values of $\phi_J$.
There is a drastic reduction of
$T$ in a wide range around $\omega/J_0 \simeq 16$.
Very narrow resonances in the region $\omega/J_0 \alt 10$
are not shown. The vertical red dotted lines and arrows show the positions 
of the resonances evaluated from Eq.\ (\ref{eq_res}) 
using the energy eigenvalues
of the time-independent Hamiltonian.
This figure is adopted from Ref.\ \cite{catform}.
}
\end{center}
\end{figure}

\begin{figure}
\begin{center}
\rotatebox{0}{
\resizebox{8.2cm}{!}
{\includegraphics{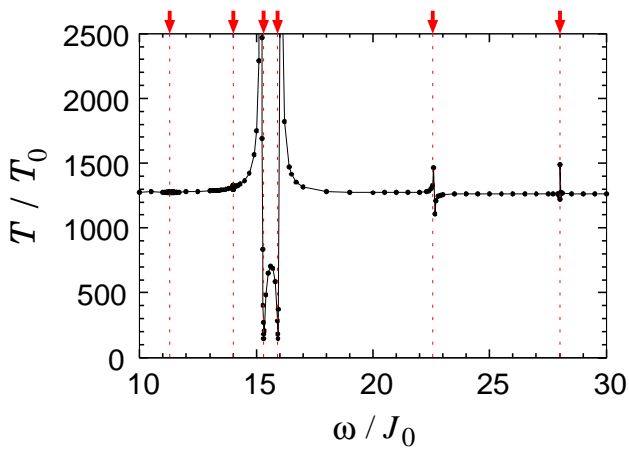}}}
\caption{\label{fig_modv}(Color online)
Tunneling period $T$ 
in the case of modulated tilt for $N = 5$, $U_0/J_0 = 4$, 
and $V_1/J_0 = 0.2$ (and $V_0=A_J=U_1=0$). 
We have set $\phi_V=0$, but there are 
no noticeable differences for different values of $\phi_V$.
The vertical red dotted lines and arrows show the positions 
of the resonances evaluated from Eq.\ (\ref{eq_res}) 
using the energy eigenvalues
of the time-independent Hamiltonian.
}
\end{center}
\end{figure}

In this section, we consider the tunneling of bosons in the
self-trapping regime characterized by $U_0N/2 J_0\gg 1$. Assume that
in the initial state all $N$ particles are either in site 1 or site 2.
The reduction of the interaction energy by single-particle tunneling
is of order $\sim U_0N$. This reduction cannot be compensated by the
increase of the kinetic energy, which is approximately given by $\sim
J_0$. As a consequence, single-particle tunneling is suppressed
(self-trapping), and all $N$ particles stay in the same well for a long
time.  In this situation, oscillations between the states $|N\rangle$
and $|-N\rangle$ occur through higher-order co-tunneling
\cite{vavtunnel}.  In Ref.\ \cite{catform} it was found that the
tunneling period of the higher-order co-tunneling can be drastically
reduced by modulating the tunneling matrix element $J$.  As we show
here, a similar phenomenon can be seen when the tilt is modulated (we
set $V_0=0$).  In Figs.\ \ref{fig_modj} and \ref{fig_modv}, we show
the tunneling period $T$ for the modulated tunneling matrix element and
tilt, respectively. As an example, we have set $N=5$ and $U_0/J_0=4$ in the both
cases.  We see that the behavior of $T$ as a function of the
modulation frequency $\omega$ depends strongly on whether $J$ or $V$
is modulated.  This difference can be explained using Floquet
analysis.


\begin{figure*}
\begin{center}
\rotatebox{0}{
\resizebox{14cm}{!}
{\includegraphics{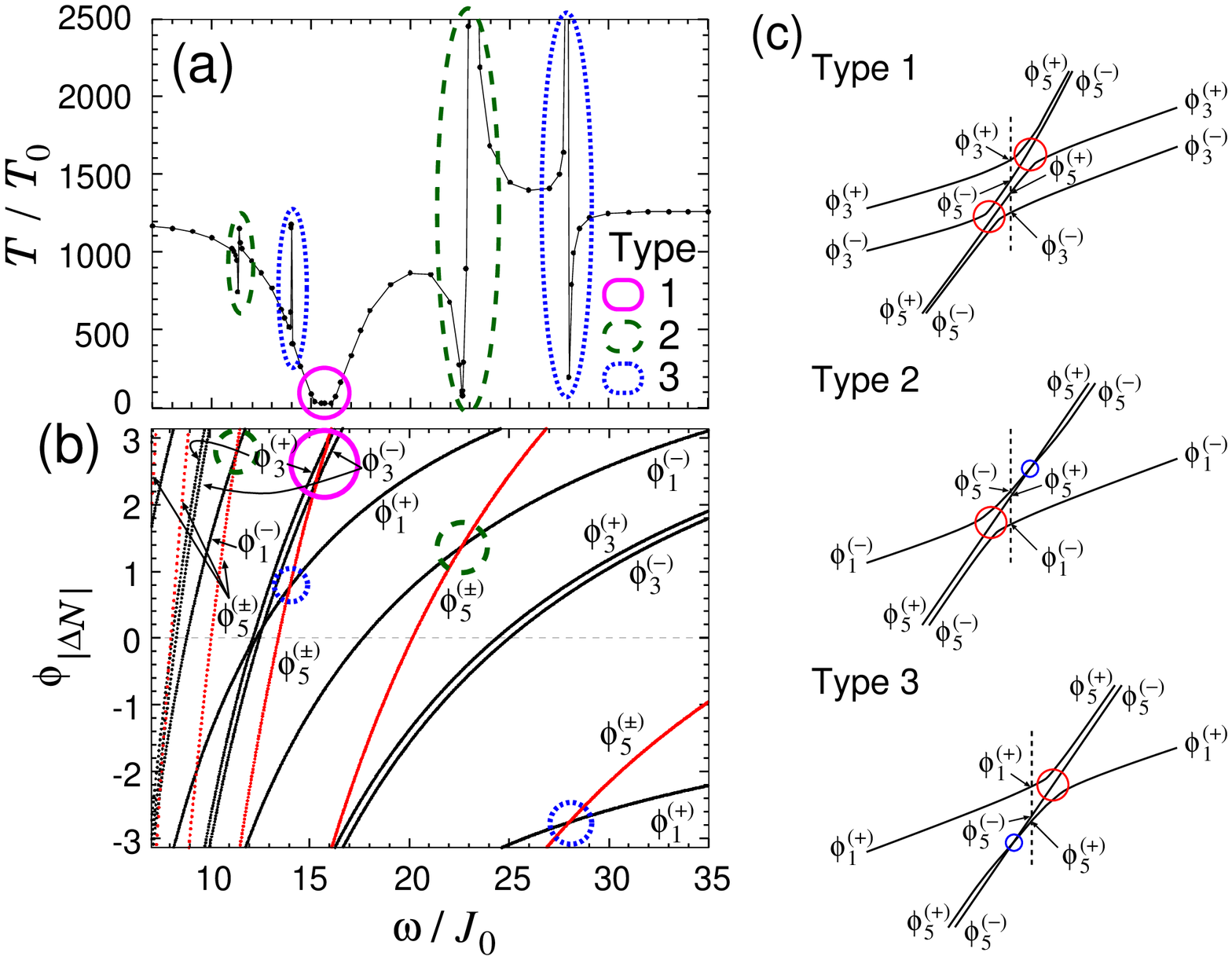}}}
\caption{\label{fig_floq}(Color online) 
Results of the Floquet analysis for the case of Fig.\ \ref{fig_modj}, 
i.e., modulated hopping parameter $J$.
Panel (a), which is the same as Fig.\ \ref{fig_modj}
(apart from the range of the horizontal axis), 
shows the tunneling period $T$ as a function of the
modulation frequency $\omega$.  Panel (b) shows the phases of the Floquet eigenvalues, 
$\phi_{|\Delta N|}$, as a function of $\omega$. 
Panel (c) shows the schematic behavior of the Floquet
eigenvalues near the crossing points in panel (b).
Each resonance observed in panel (a) corresponds to 
one of the three types of crossings shown in panel (c).
In panels (a) and (b), crossings of types 1, 2, and 3 
are labeled by the magenta solid,
green dashed, and blue dotted curves, respectively.
}
\end{center}
\end{figure*}

Before analyzing the system in detail, we first summarize two key
points.  One is the parity of the operator whose coefficient is
modulated, and the other is the shift in the phases of the Floquet
eigenvalues due to an avoided crossing.  The parity of $\hat{S}_x$ is
even and that of $\hat{S}_z$ is odd.  Therefore, $\hat{S}_x$ couples
Floquet eigenstates of the same parity, and $\hat{S}_z$ couples those
of the opposite parity.  This means that in the case of modulated $J$
[$V$], there is an avoided crossing between Floquet eigenstates of the
same [opposite] parity. The differences in the behavior of the
tunneling period can be attributed to the parities of the states
undergoing an avoided crossing.  Below we show that usually
$\phi_i^{(+)} > \phi_i^{(-)}$ ($\phi_i^{(-)} > \phi_i^{(+)}$) holds
for odd (even) $N$.  However, this is not necessarily the case near
avoided crossings where the values of $\phi_i^{(\pm)}$ are
shifted. These shifts lead to either suppression or enhancement of the
tunneling.

We have chosen $N=5$ and $U_0/J_0=4$ in this section.  The results
can, however, be straightforwardly generalized to any value of $N$ and
$J_0/U_0$, as long as $N>1$ and $U_0 N/J_0 \gg 1$.  We remark that
Figs.\ \ref{fig_modj}, \ref{fig_floq}(b), and the top panel of
Fig.\ \ref{fig_amp} are taken from Ref.~\cite{catform}, but the
detailed Floquet analysis of the $J$ modulation, as well as the entire
analysis of the effects of the tilt and interaction modulation, has
not been presented elsewhere.

\subsection{Time-independent Hamiltonian}\label{sec_timeindep}

If the modulation amplitude is small and the system is not near an avoided crossing, the Floquet eigenstates and eigenvalues turn out to be close to the ones determined by the time-independent part of the Hamiltonian, given by 
\begin{align}
\label{H0}
\hat{H}_0 = -2 J_0\hat{S}_x + U_0 \hat{S}_z^2.
\end{align}
As a consequence, some important properties of the modulated system, such as the positions of the resonances, can be explained by analyzing the spectrum of $\hat{H}_0$.

We assume that $U_0N \gg J_0$ and $V_0=0$.  In order to compare the
time-evolution operator of the original time-dependent modulated
system with that determined by Hamiltonian (\ref{H0}), we define
the Floquet operator $\hat{F}_0$ corresponding to $\hat{H}_0$ as
\begin{equation}
\label{eq_F_nonmodulated}
\hat{F}_0=e^{-i T_\omega \hat{H}_0},\quad T_\omega=\frac{2\pi}{\omega}. 
\end{equation} 
We denote the eigenvalues of the time-independent Hamiltonian by
$E_{0;k}^{(\pm)}$, where we use the same indexing as in the case of
the eigenvectors of the Floquet operator $\hat{F}$.  The phases of the
Floquet eigenvalues are given by
\begin{align}
\label{eq_phi_nonmodulated}
  \phi_{0;k}^{(\pm)} = -E_{0;k}^{(\pm)} T_\omega \quad \textrm{mod } 2\pi.
\end{align}
We find that $E_{0;i}^{(+)} < E_{0;i}^{(-)}$ for odd $N$ and
$E_{0;i}^{(+)} > E_{0;i}^{(-)}$ for even $N$.  Because of the minus
sign in Eq.\ $(\ref{eq_phi_nonmodulated})$, the opposite holds for the
phases of the Floquet eigenvalues $\phi_{0;k}^{(\pm)}$. The situation
is similar in the presence of a small-amplitude modulation, and thus,
normally, $\phi_i^{(+)} > \phi_i^{(-)}$ ($\phi_i^{(-)} > \phi_i^{(+)}$)
for odd (even) $N$.  Now $E_{0;k} > E_{0;l}$ if $k>l\geq 0$. Using
this and the equation $\partial_\omega
\phi_{0;k}^{(\pm)}=E_{0;k}^{(\pm)} 2\pi/\omega^2$, we see that
$\phi_{0;k}^{(\pm)}$, and therefore also $\phi_{k}^{(\pm)}$, increases
faster as a function of $\omega$ the larger $k$ is.  This means
that if $k>l$, $\phi_{k}$ approaches $\phi_{l}$ from below as $\omega$
increases [see Fig.\ \ref{fig_floq}(b) for an example]. The phases
$\{\phi_k^{(\pm)}\}$ cross repeatedly as $\omega$ increases.  A
crossing occurs when $\omega$ satisfies
\begin{equation}
  n\omega \approx |E_{0;k}-E_{0;l}|\, , \label{eq_res}
\end{equation}
with $n=1,2,3,\ldots$.
The last crossing between $\phi_k$ and $\phi_l$ is
at $\omega \approx |E_k-E_l|$.
In the limit of $\omega\rightarrow\infty$,
the phases of all the Floquet eigenvalues approach zero from the negative side.

In the specific case $N=5$ and $U_0/J_0=4$, corresponding to Figs.\ \ref{fig_modj} and \ref{fig_modv},
the crossing points between $\phi_5$ and the other phases   
are at $\omega/J_0=28.01$ (crossing with $\phi_1^{(+)}$),
$22.63$ ($\phi_1^{(-)}$), $15.93$ ($\phi_3^{(+)}$), and
$15.31$ ($\phi_3^{(-)}$).  
These are obtained from Eq.\ (\ref{eq_res}) with $n=1$. 
Note that $E_5^{(+)}$ and $E_5^{(-)}$, and  
thus $\phi_5^{(+)}$ and $\phi_5^{(-)}$, are almost identical.

The crossing points corresponding to 
$n=1$ and $2$ in the region $\omega/J_0>10$ are shown
by the vertical red dotted lines and arrows 
in Figs.\ \ref{fig_modj} and \ref{fig_modv}.
We see that the positions of all the resonances shown in Figs.\ \ref{fig_modj} and \ref{fig_modv} are
well explained by the energy eigenvalues of the time-independent Hamiltonian.
Based on this fact, we can say that the positions of the crossings
are the same irrespective of the modulated variable.

\subsection{Modulated $J$}

Next we consider the modulation of the tunneling matrix element; 
see Figs.\ \ref{fig_modj} and \ref{fig_floq}. 
The value $\langle\psi_i| \hat{S}_x|\psi_j\rangle$ can be non-zero only 
if the Floquet eigenstates $\psi_i$ and $\psi_j$ have the same parity. 
Consequently, there is an avoided crossing between eigenstates with the same parity.

In Fig.\ \ref{fig_floq}(b), we show 
the phases of the Floquet eigenvalues as a function of $\omega /J_0$
for the parameters used in Fig.\ \ref{fig_modj} 
[and Fig.\ \ref{fig_floq}(a)].   
From Fig.\ \ref{fig_floq} we see that a large change of $T$ occurs when   
 $\phi_5^{(\pm)}$ crosses the other $\phi_{|\Delta N|}^{(\pm)}$'s  
[circles and ellipses in Figs.\ \ref{fig_floq}(a) and \ref{fig_floq}(b)].
The behavior of the phases of the Floquet eigenvalues near the crossings is schematically shown in Fig.\ \ref{fig_floq}(c).
In an $N$-particle system, there are $N-2$ different types of crossings 
between $\phi_{|\Delta N|=N}^{(\pm)}$ and other  $\phi_{i}^{(\pm)}$'s \cite{note_crossings}. 
Because now $N=5$, we have three types of crossings;
each resonance corresponds to one of these. 
In the following, we analyze in detail each of these three crossing types.

\subsubsection{Type 1}

A crossing between $\phi_5^{(\pm)}$ and $\phi_3^{(\pm)}$ 
leads to a reduction of $T$ in a wide range of $\omega$
around $\omega/J_0 \simeq 16$. This crossing is indicated 
in Figs.\ \ref{fig_floq}(a) and \ref{fig_floq}(b) by  the solid magenta circle. 
The detailed structure of the crossing is shown schematically in the top figure 
in Fig.\ \ref{fig_floq}(c).  
Since $\phi_3^{(+)}$ and $\phi_3^{(-)}$ are almost equal, 
the avoided crossings between $\phi_3^{(-)}$ and $\phi_5^{(-)}$ and 
between $\phi_3^{(+)}$ and $\phi_5^{(+)}$ occur almost simultaneously 
[the red solid circles in Fig.\ \ref{fig_floq}(c)].
Because $\phi_i^{(+)} > \phi_i^{(-)}$ 
for odd $N$ outside the crossing region (see Sec.\ \ref{sec_timeindep}) 
and $\hat{S}_x$ couples Floquet eigenstates with the same parity, 
 the first avoided crossing occurs between $\phi_3^{(-)}$ and $\phi_5^{(-)}$
[the left red solid circle] as the modulation frequency increases.
Due to the repulsion between these two levels,  
the splitting between $\phi_5^{(\pm)}$ is increased near the avoided crossing and thus the tunneling period is reduced.
The second avoided crossing takes place between $\phi_3^{(+)}$ and $\phi_5^{(+)}$ [the right red solid circle].
Note that, after the first avoided crossing, 
the states $\psi_3^{(-)}$ and $\psi_5^{(-)}$ have been interchanged 
[between the red solid circles]
and the energy splitting between $\phi_5^{(\pm)}$ remains large
until the second avoided crossing 
at which $\psi_3^{(+)}$ and 
$\psi_5^{(+)}$ are interchanged.
These successive avoided crossings lead to a 
reduction of the tunneling period in a wide range
of $\omega/J_0$.

\subsubsection{Type 2}

Because of the large quasienergy splitting between $\phi_1^{(+)}$ and
$\phi_1^{(-)}$, the points where $\phi_5^{(\pm)}$ crosses
$\phi_1^{(+)}$ and $\phi_1^{(-)}$ are far apart.  We call a crossing
between $\phi_5^{(\pm)}$ and $\phi_1^{(-)}$ a type 2 crossing and that
between $\phi_5^{(\pm)}$ and $\phi_1^{(+)}$ a type 3 crossing.  With
increasing $\omega$, a type 2 crossing first yields a reduction and
then an enhancement of the tunneling period.  We show the schematic
structure of a type 2 crossing in the middle figure in
Fig.\ \ref{fig_floq}(c).  The resonances around $\omega/J_0\simeq 11$
and $\omega/J_0\simeq 23$, indicated by the green dashed curves in
Figs.\ \ref{fig_floq}(a) and \ref{fig_floq}(b), correspond to type 2
crossings.

Suppose that the crossing is approached from the small $\omega/J_0$ side.
Far from the avoided crossing $\phi_5^{(+)} > \phi_5^{(-)}$ as 
explained in Sec.\ \ref{sec_timeindep}.
Since $\hat{S}_x$ couples Floquet eigenstates with the same parity,
$\phi_1^{(-)}$ undergoes an avoided crossing with $\phi_5^{(-)}$
(the large red solid circle). 
Near the avoided crossing, the energy splitting between $\phi_5^{(\pm)}$
is enhanced, which leads to the reduction of the tunneling period.
Just after the avoided crossing (around the vertical dashed line), 
the states $\psi_1^{(-)}$ and $\psi_5^{(-)}$ are interchanged and, 
unlike in the usual situation, $\phi_5^{(-)} > \phi_5^{(+)}$.
Since $\phi_5^{(+)}$ is larger than $\phi_5^{(-)}$ 
far from the avoided crossing also on the large $\omega/J_0$ side, 
$\phi_5^{(+)}$ and $\phi_5^{(-)}$ cross each other 
(the small blue solid circle), 
which yields a divergence of the tunneling period.

\subsubsection{Type 3}

As opposed to a type 2 crossing, a type 3 crossing (crossings between $\phi_5^{(\pm)}$ and $\phi_1^{(+)}$) gives first an enhancement and then a reduction of the tunneling period.
The resonances at $\omega/J_0\simeq 14$ and $\omega/J_0\simeq 28$, 
which are indicated by the blue dotted ellipses and circles in Figs.~\ref{fig_floq}(a) and \ref{fig_floq}(b), 
correspond to type 3 crossings. 
A detailed schematic structure of a type 3 crossing is shown
in the bottom figure in Fig.\ \ref{fig_floq}(c).
Suppose again that we approach the crossing from the small $\omega/J_0$ side.
In this case, we have an avoided crossing between 
$\phi_1^{(+)}$ and $\phi_5^{(+)}$.
The phase $\phi_5^{(+)}$, which is located above
$\phi_5^{(-)}$ far from the crossing, is pushed downward 
due to the avoided crossing with $\phi_1^{(+)}$, and thus
$\phi_5^{(+)}$ and $\phi_5^{(-)}$ cross each other 
(the small blue solid circle).  This leads to the divergence of the
tunneling period.  After this, there is an avoided crossing
between $\phi_1^{(+)}$ and $\phi_5^{(+)}$ (the large red solid circle),
leading to an enhancement of the quasienergy splitting between $\phi_5^{(\pm)}$. 
This yields a reduction in the tunneling period.
After the avoided crossing, the states $\psi_1^{(\pm)}$ and $\psi_5^{(\pm)}$ are interchanged.

\subsubsection{Type 1$'$: Small $A_J$}

\begin{figure}
\begin{center}
\rotatebox{0}{
\resizebox{7.5cm}{!}
{\includegraphics{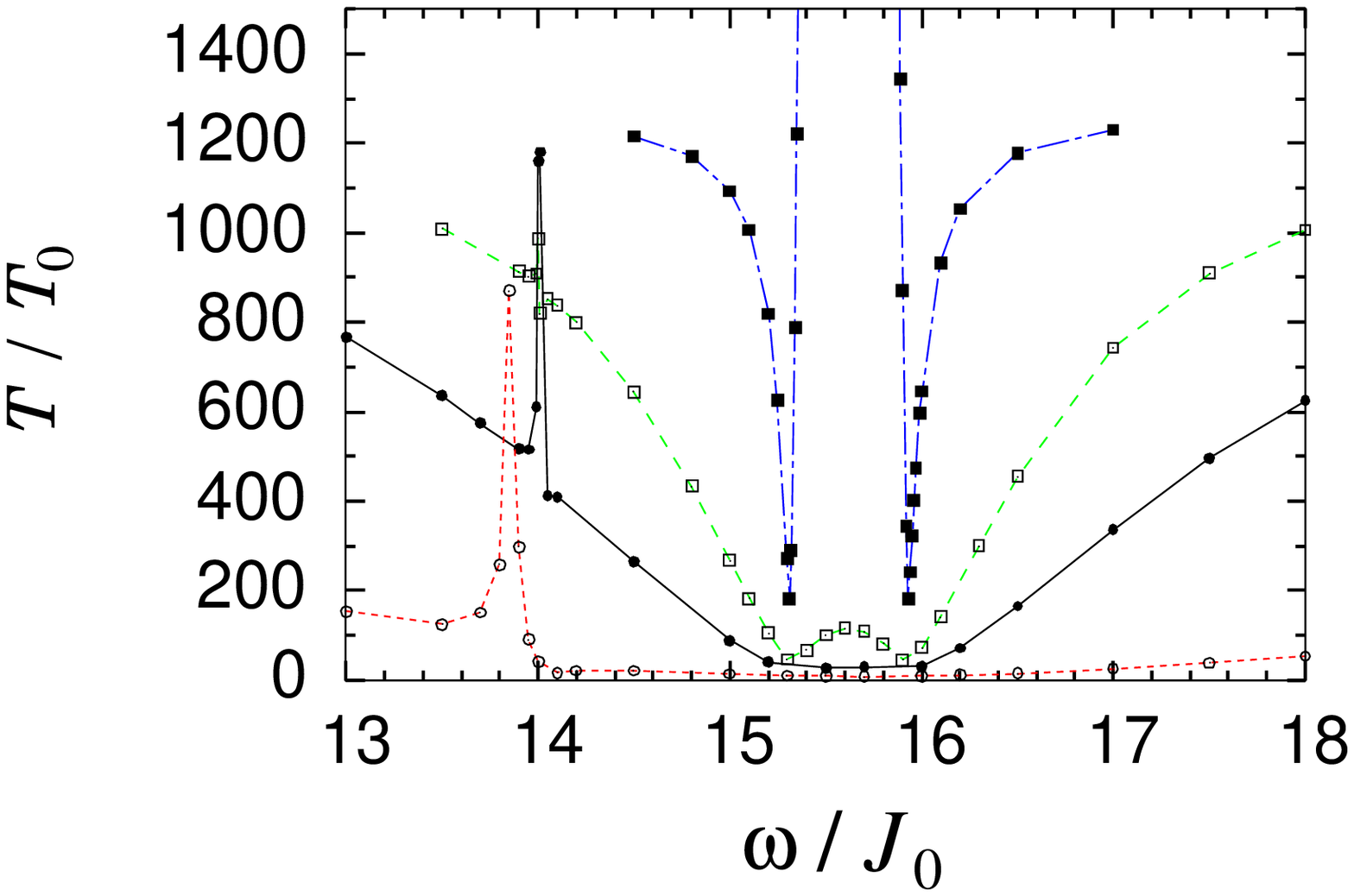}}}
\rotatebox{0}{
\resizebox{7.2cm}{!}
{\includegraphics{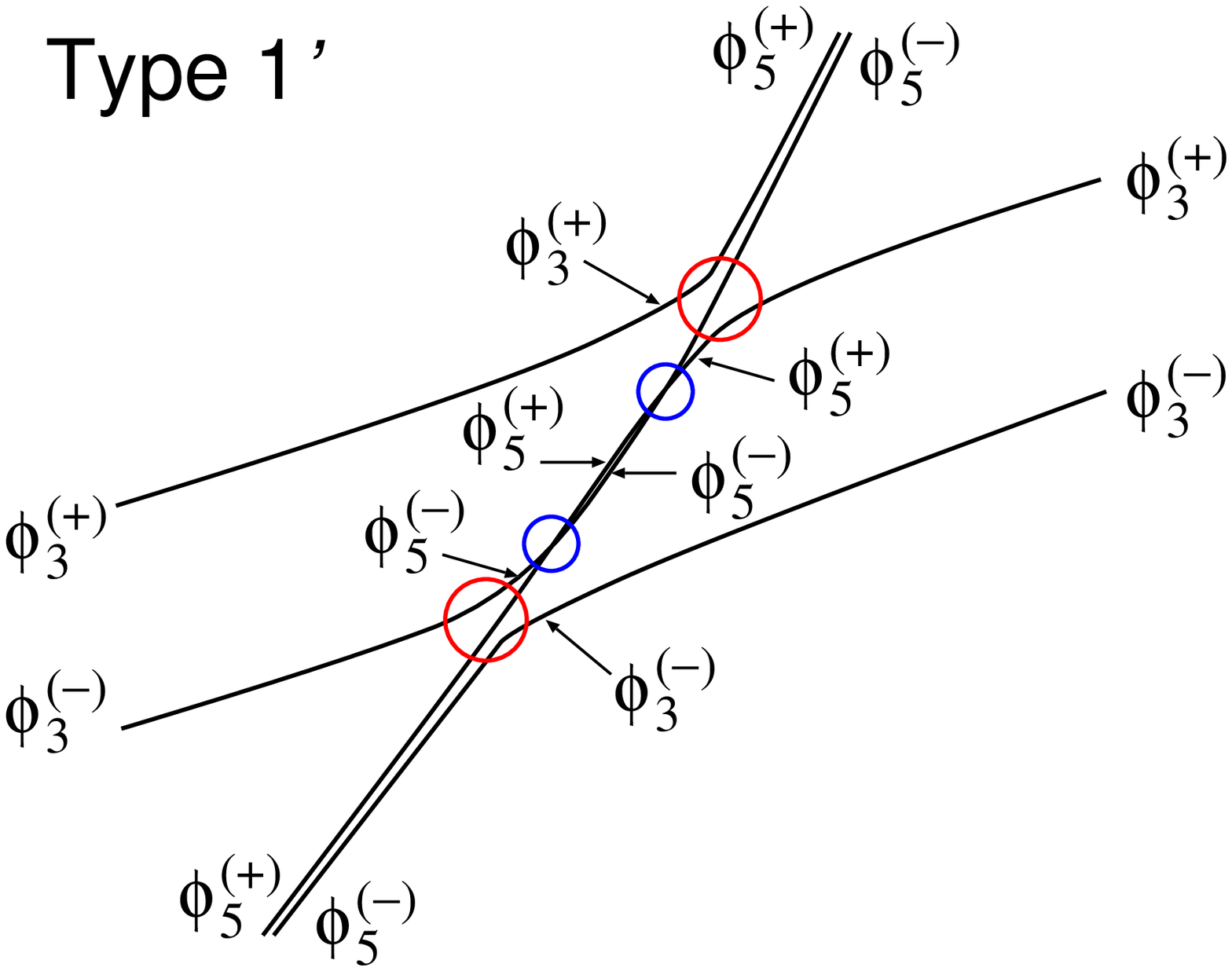}}}
\caption{\label{fig_amp}(Color online) 
The top panel (taken from Ref.\ \cite{catform})
shows the tunneling period $T$ in the case of modulated tunneling matrix element $J$ 
for various values of the modulation amplitude. 
The amplitudes used are 
$A_J= 0.5$ (red dotted line), $0.1$ (black solid line),
$0.05$ (green dashed line), and $0.01$ (blue dashed-dotted line).
The other parameters are the same as in Figs.\ \ref{fig_modj} and \ref{fig_floq}:
$N=5$ and $U_0/J_0=4$ (and $V_0=V_1=0$).  
The tunneling period does not depend on $\phi_J$ noticeably,
and here we have set $\phi_J=0$ for definiteness. \quad
The bottom panel shows the schematic behavior of 
the phases of the Floquet eigenvalues near the crossing between $\phi_5^{(\pm)}$
and $\phi_3^{(\pm)}$ for small values of $A_J$ 
(corresponding to, e.g., $A_J=0.01$ in the top panel)
compared to the type 1 case shown in Fig.\ \ref{fig_floq}(c). 
We call this a type 1$'$ crossing.
}
\end{center}
\end{figure}

In the top panel of Fig.\ \ref{fig_amp}, we show the tunneling
period $T$ for various values of the modulation amplitude $A_J$.  With
decreasing $A_J$, the resonance around $\omega/J_0\simeq 16$ becomes
narrower and finally separates into two resonances (see the case
$A_J=0.01$ shown by the blue dashed-dotted line).  The schematic
behavior of the phases of the Floquet eigenvalues near the crossing is
shown in the bottom panel of Fig.\ \ref{fig_amp}.  We call this a type
1$'$ crossing.  The major difference between type 1$'$ and type 1
crossings is the existence of two points where $\phi_5^{(\pm)}$ cross
each other. These are indicated by the small blue solid circles, and
they are located between the two avoided crossings (the large red
solid circles).  One can also view a type 1$'$ crossing as a
combination of type 2 and type 3 crossings.

When $A_J$ is small, the coupling between the two states that undergo
an avoided crossing is small.  Thus the difference 
$|\phi_5^{(+)}-\phi_5^{(-)}|$ remains very small even near the avoided crossing.
Therefore, unlike in a type 1 crossing, 
the inverted configuration of $\phi_5^{(\pm)}$ (i.e., the situation
$\phi_5^{(-)}>\phi_5^{(+)}$) cannot be sustained 
throughout the region between the two avoided
crossings. This leads to the appearance of 
two crossing points, indicating a diverging tunneling period.
In Ref.\ \cite{catform} it has been shown that the divergences are present if  
$A_J \lesssim N^{-1} (J_0/U_0)^{N-3} (N-1)(N-2)/(N-3)!$.

\subsection{Modulated $V$}\label{sec_modv}

\begin{figure}
\begin{center}
\rotatebox{0}{
\resizebox{7.2cm}{!}
{\includegraphics{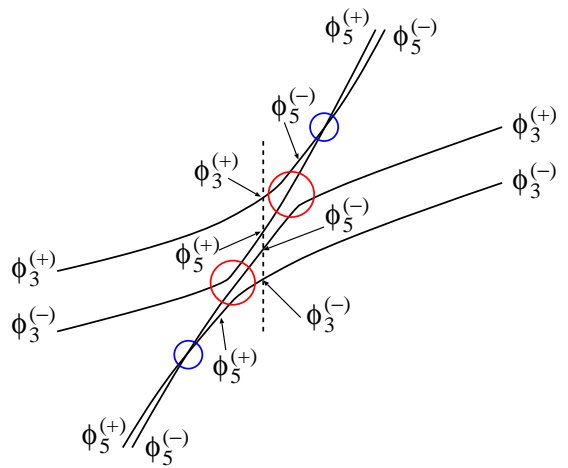}}}
\caption{\label{fig_floq_tilt}(Color online) 
Schematic behavior of the phases of the Floquet eigenvalues
near the resonance around $\omega/J_0\simeq 16$ 
in the case of modulated $V$ shown in Fig.\ \ref{fig_modv}.
}
\end{center}
\end{figure}

As can be seen from Figs. \ref{fig_modj} and \ref{fig_modv}, the
tunneling period behaves differently when $V$ is modulated.  In
Fig.~\ref{fig_modv}, a noticeable change in the tunneling period $T$
can be seen around $\omega/J_0\simeq 16$. There are also small, narrow
resonances at $\omega/J_0 \simeq 22.5$ and $\omega/J_0 \simeq 28$
\cite{note_narrowres}.  Unlike in the case of modulated $J$, the
resonance at $\omega/J_0\simeq 16$ is not a wide and smooth reduction
of $T$ for any value of the modulation amplitude $V_1$.

As in the case of modulated $J$, the resonance around
$\omega/J_0\simeq 16$ is caused by a crossing between $\phi_5^{(\pm)}$
and $\phi_3^{(\pm)}$.  However, unlike $\hat{S}_x$, the operator
$\hat{S}_z$ has odd parity, and it thus couples Floquet eigenstates of
opposite parity.  In Fig.\ \ref{fig_floq_tilt}, we show the schematic
behavior of the Floquet eigenstates near $\omega/J_0\simeq 16$.
Suppose that the crossing is approached from the small $\omega/J_0$
side.  As $\omega/J_0$ increases, the states $\psi_3^{(-)}$ and
$\psi_5^{(+)}$ undergo an avoided crossing, and $\phi_5^{(+)}$ is
pushed downward.  Far from the avoided crossing the relation
$\phi_i^{(+)} > \phi_i^{(-)}$ holds.  Because of this and the fact
that $\phi_5^{(+)}$ is pushed downward, the phases $\phi_5^{(+)}$ and
$\phi_5^{(-)}$ cross (the left small blue circle) before the avoided
crossing (the left large red circle).  After the first avoided
crossing, the states $\psi_3^{(-)}$ and $\psi_5^{(+)}$ are
interchanged. Next, $\phi_3^{(+)}$ and $\phi_5^{(-)}$ undergo an
avoided crossing (the right large red circle), and the corresponding
states are interchanged.  Because now $\phi_5^{(-)}>\phi_5^{(+)}$,
these phases cross after the second avoided crossing (the right small
blue circle), so that $\phi_5^{(+)}>\phi_5^{(-)}$ far away from the
crossing.
The two crossing points and the two avoided crossings correspond to
the two divergences and the two reductions of the tunneling period,
respectively. These are shown in Fig.\ \ref{fig_modv} near
$\omega/J_0\simeq 16$.  Because in the present case the avoided
crossings occur between Floquet eigenstates of opposite parity, the
phases $\phi_5^{(\pm)}$ necessarily cross each other outside the
region of the successive avoided crossings.  For this reason, a smooth
reduction of $T$ in a wide region of the modulation frequency $\omega$
cannot be achieved by modulating the tilt.  This is one of the major
findings of this paper.

We note that all the other resonances are also much narrower 
than in the case of modulated $J$.
This is because the operator $\hat{S}_z$, which is related to the tilt,
does not contribute to the single-particle tunneling, unlike $\hat{S}_x$.
The range of $\omega$ characterizing the width of the resonance
is comparable to the quasienergy separation at the avoided crossing.
This is approximately proportional to 
$|\langle\psi_5^{(\pm)}|\hat{S}_x|\psi_{i\ne5}^{(\mp)}\rangle|^2$
in the case of $J$ modulation and to
$|\langle\psi_5^{(\pm)}|\hat{S}_z|\psi_{i\ne5}^{(\pm)}\rangle|^2$
in the case of $V$ modulation. The latter is smaller than the former
by a factor $\sim (J_0/U_0)^2$.
This will be discussed in more detail in Sec.\ \ref{sec_noon}.

\subsection{Modulated $U$}

Finally, we consider the case in which 
the on-site interaction strength $U$ is modulated weakly $(U_1/U_0\ll 1)$.
The Hamiltonian in this case is $\hat{H}=-2J_0 \hat{S}_x + U(t) \hat{S}_z^2$,
with $U(t)$ given by Eq.\ (\ref{eq_ut}).  Since this Hamiltonian can be
rewritten as
\begin{equation}
  \hat{H}(t)= {\cal A}(t)
\left[-2J_{\rm eff}(t) \hat{S}_x + U_0 \hat{S}_z^2 \right],
\end{equation}
with ${\cal A}(t) = 1+ (U_1/U_0)\sin{(\omega t+\phi_U)}$ and
\begin{equation}
  J_{\rm eff}(t) 
\simeq J_0 \left[1+\frac{U_1}{U_0}\sin{(\omega t+\phi_U+\pi)}\right],
\end{equation}
we can expect that the dynamics can be reproduced by modulating $J$ with the amplitude 
$A_J=U_1/U_0$ and phase $\phi_J=\phi_U+\pi$ instead of modulating $U$.

\begin{figure}
\begin{center}
\rotatebox{0}{
\resizebox{8.cm}{!}
{\includegraphics{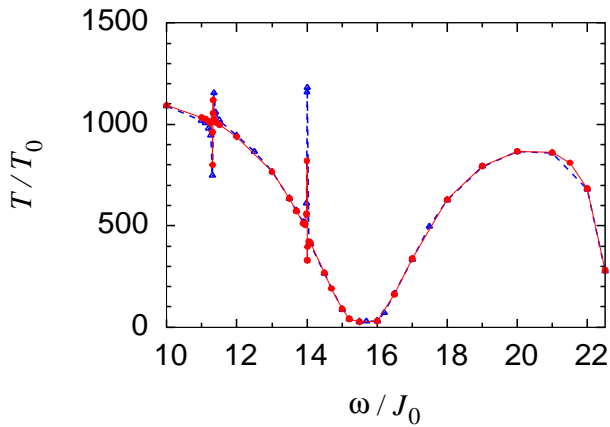}}}
\caption{\label{fig_modu}(Color online) 
Tunneling period $T$ in the case of $U$ modulation (red solid line). For comparison, the tunneling period 
corresponding to $J$ modulation is also shown (blue dashed line). Here $N=5$, $U_0/J_0=4$, and $V_0=V_1=0$. 
In the case of $J$ modulation $A_J=0.1$ and $U_1=0$, while in the case 
of $U$ modulation $A_J=0$ and $U_1/J_0 =0.4$.}
\end{center}
\end{figure}

This observation is confirmed by the result shown in
Fig.\ \ref{fig_modu}, where we compare the tunneling period $T$ as a
function of $\omega$ in the cases of modulated $J$ and modulated
$U$. In this example $N = 5$ and $U_0/J_0 = 4$.  The result for the
modulated $J$ is taken from Fig.\ \ref{fig_modj} ($A_J = 0.1$).  By
setting $U_1=A_JU_0$, i.e., $U_1/J_0=A_J U_0/J_0=0.4$ in the present
case, these two results almost coincide with each other. Note that
since $T$ does not noticeably depend on the phase of the modulation,
$U_1=A_JU_0$ is a sufficient condition for the behaviors of the
tunneling periods to coincide.

An analysis of the phases of the Floquet eigenvalues in the case of
the $U$ modulation shows that the schematic behavior of these phases
around each resonance is the same as in the case of the $J$ modulation
shown in Fig.\ \ref{fig_floq}(c).  This can be understood by noting
that $\hat{S}_x$ and $\hat{S}_z^2$ have the same parity.

\section{Coherent destruction of tunneling and the Floquet spectrum}\label{sec_Josephson}

In this section, we first study a system characterized by a weak
interaction and a large-amplitude tilt modulation, concentrating on
the properties of the Floquet spectrum.  After this we examine the
effects of a large-amplitude modulation of the interaction.  The
Floquet spectrum of this system has been analyzed elsewhere (see
Refs.\ \cite{strzys08,gong09}) and will not be discussed
here. Instead, we propose a way to create NOON states using selective
tunneling originating from the modulation of the interaction strength.

\subsection{Modulated $V$}

\begin{figure}
\begin{center}
\rotatebox{0}{
\resizebox{8cm}{!}
{\includegraphics{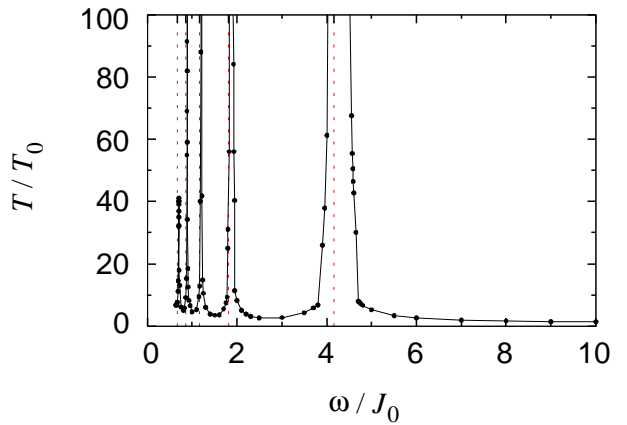}}}
\caption{\label{fig_cdt}(Color online)
Tunneling period in the weak interaction regime with
large-amplitude modulation of the tilt.
The parameters are $N=5$, $U_0/J_0=0.1$, $V_0=0$, and $V_1/J_0=10$
(and $A_J=U_1=0$).
We have set $\phi_V=0$ in this calculation, but 
the behavior of $T$ does not depend noticeably on $\phi_V$.   
The vertical red dotted lines correspond to the values of $\omega/J_0$
which give ${\cal J}_0(V_1/\omega)=0$.
}
\end{center}
\end{figure}

Next, we consider a case where the interaction is weak, $UN/J_0\alt
1$, and the amplitude of the modulation of the tilt is large,
$V_1/J_0\gg 1$. We assume that the tunneling matrix element $J$ and
the interaction strength $U$ are time-independent, that is, $A_J=0$
and $U_1=0$. Furthermore, we set $V_0=0$. In this case, it is
well-known that the effect of the modulation of the tilt can be
approximately described by a renormalized tunneling term. In more
detail, the original tunneling term $\hat{T} \equiv -2 J \hat{S}_x$ is
replaced by an effective one
\cite{dunlap,holthaus,haroutyunyan,eckardt,creffield08,luo08,kudo}:
\begin{align}
\nonumber 
\hat{T}_{\rm eff} =& -2J_0 {\cal J}_0\left(\frac{V_1}{\omega}\right)
\Big\{ \cos{\left[\frac{V_1}{\omega}\cos{\phi_V}\right]} \hat{S}_x\\ 
& - \sin{\left[\frac{V_1}{\omega}\cos{\phi_V}\right]} \hat{S}_y \Big\},
\end{align}
where ${\cal J}_0$ is the zeroth-order Bessel function 
(see Appendix \ref{sect_jeff} for the derivation).
Coherent destruction of tunneling takes place  
when $V_1/\omega$ is equal to one of the zeros of ${\cal J}_0$.
In the rest of this section, we discuss CDT in terms of the Floquet eigenvalues. 
This discussion holds for any $N\ge1$.

In Fig.\ \ref{fig_cdt}, we show the tunneling period $T$ as a function of
the modulation frequency $\omega$ in the regime of weak interaction and large-amplitude modulation. In this calculation, we have set
$N=5$, $U_0/J_0=0.1$, $V_0=0$, $V_1/J_0=10$, and $\phi_V=0$,
and in the initial state  all particles
are in site 1. The first five zeros of ${\cal J}_0(V_1/\omega)$ are at
$V_1/\omega= 2.40$, $5.52$, $8.65$, $11.79$, and $14.93$:
they correspond to $\omega/J_0= 4.16$, $1.81$, $1.16$, $0.848$, and $0.670$,
respectively. These frequencies are shown by the vertical red dotted lines
in Fig.\ \ref{fig_cdt}.
There is good agreement between these dotted lines and the actual positions
of the peaks of $T$.

\begin{figure}
\begin{center}
\rotatebox{0}{
\resizebox{8cm}{!}
{\includegraphics{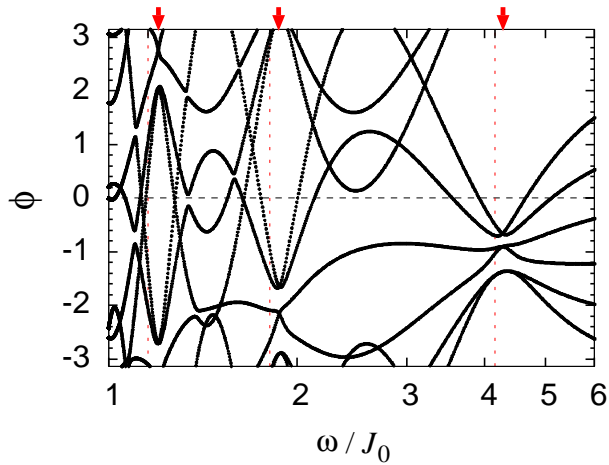}}}
\caption{\label{fig_floq_cdt}(Color online)
Phases of the Floquet eigenvalues in the case of Fig.\ \ref{fig_cdt}.
The parameters are $N=5$, $U_0/J_0=0.1$, $V_0=0$, and $V_1/J_0=10$ 
(and $A_J= U_1 =0$).
In this calculation we have set $\phi_V=0$, but the phases of the Floquet eigenvalues 
do not depend on $\phi_V$.
The vertical red dotted lines correspond to the values of $\omega/J_0$
which give ${\cal J}_0(V_1/\omega)=0$.
The red arrows show the actual positions of the peaks of $T$ (see Fig.\ \ref{fig_cdt}).
}
\end{center}
\end{figure}

In Fig.\ \ref{fig_floq_cdt}, we plot the phases of the Floquet
eigenvalues for the parameters used in Fig.\ \ref{fig_cdt}. When the
CDT occurs, the phases gather in pairs, the phases in each pair being
almost equal, and all the pairs gather in a narrow region (red arrows
in Fig.\ \ref{fig_floq_cdt}).  This behavior can be understood by
noting that the Hamiltonian is effectively $\simeq U_0 \hat{S}_z^2$ at
the point where CDT occurs, and thus $\Delta N$ becomes a good quantum
number, with a twofold degeneracy with respect to $\pm\Delta N$.

Finally, we discuss the difference between even and odd $N$ cases.
For even $N$, the number of the Floquet eigenvalues is $N+1$, which is odd.
Therefore, when CDT occurs, the Floquet eigenvalues are grouped into one trio
and $(N-2)/2$ pairs [cf. $(N+1)/2$ pairs for odd $N$].
A key point is that, for even $N$, there is a Fock state $|\Delta N =0\rangle$,
which does not have a degenerate pair, unlike the other Fock states. 
In this case, the Floquet eigenstates near the value of $\omega$ at which CDT occurs 
can be classified into three types: 1) one Floquet eigenstate 
that has maximum amplitude at $\Delta N=0$ component,
2) $N/2$ Floquet eigenstates that do not have maximum amplitude at the 
$\Delta N=0$ component but that always have a non-zero $\Delta N=0$ component, and
3) $N/2$ Floquet eigenstates that do not have maximum amplitude at the 
$\Delta N=0$ component and where this component becomes zero
when CDT occurs.
The trio consists of all the three types, and the $(N-2)/2$ pairs
consist of the second and third types.
We note that, for even $N$, the degeneracies of the trio and of all the pairs
are incomplete provided $U_0\ne 0$ \cite{note_u0=0}, 
while all the pairwise degeneracies
are complete for odd $N$.  Consequently, CDT is more complete for odd $N$
than even $N$.

\subsection{Modulated $U$}

Due to the non-linear dependence of the interaction on $\Delta N$,
the CDT caused by a large-amplitude modulation of the interaction strength
($U_1\gg J_0$, $U_0$) is state dependent \cite{gong09}.
Here we assume $A_J=V=0$ for simplicity.
In this case, a condition for partial CDT between the states
$|\Delta N=m\rangle$ and $|\Delta N=m-2\rangle$ 
($m$ is a positive integer) is
\begin{equation}
  {\cal J}_0\left[\frac{U_1}{\omega} (m-1)\right] =0;
\end{equation}
see Appendix \ref{sect_umod} for the derivation.

Unlike in the case of modulated $V$ shown in Fig.\ \ref{fig_floq_cdt},
only the Floquet eigenstates relevant to partial CDT show the
degeneracy in the phases of the Floquet eigenvalues (see, e.g., Fig.~1
of Ref.\ \cite{gong09}).  For odd $N$, each partial CDT is associated
with a perfect degeneracy of the phases of the Floquet eigenvalues
while, for even $N$, some degeneracies (but not all) are incomplete
provided $U_0\ne 0$.  As in the case of modulated $V$, these
incomplete degeneracies are caused by the existence of the Fock state
$|\Delta N=0\rangle$.  Consequently, partial CDT is generally more
complete for odd $N$ than for even $N$.  The Floquet spectrum in the
case of large-amplitude modulation of $U$ and weak interaction has
been studied in depth in Refs.\ \cite{strzys08,gong09}.  We refer to
these references for further discussion.

\begin{figure}[tb]
\begin{center}
\rotatebox{270}{
\resizebox{!}{8.5cm}
{\includegraphics{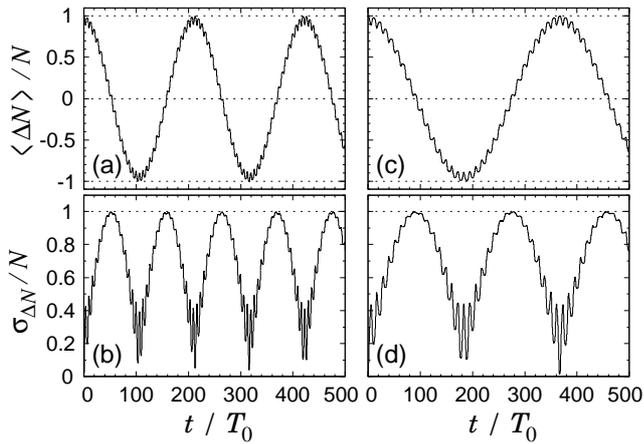}}}
\caption{\label{fig_noonmodu} 
Time evolution of the normalized population imbalance 
$\langle \Delta N\rangle/N$ and its variance 
$\sigma_{\Delta N}/N\equiv N^{-1}\sqrt{\langle \Delta N^2\rangle -\langle \Delta N\rangle^2}$ 
under large-amplitude modulation of $U$. Here 
 $N=21$ [panels (a) and (b)] and $N=51$ [panels (c) and (d)], and the initial state is $|\Delta N= N\rangle$.  
In the case $N=21$ we have set $U_1/J_0=10$ and $\omega/J_0=83.85$, and 
in the case $N=51$ we have set $U_1/J_0=4$ and $\omega/J_0=83.4$.
Other parameters are $U_0=J_1=V_0=V_1=0$.
A coherent oscillation between $|N\rangle$ and $|-N\rangle$ is realized 
by slightly detuning from a partial CDT between 
Floquet eigenstates $\psi_N^{(\pm)}$.
}
\end{center}
\end{figure}

Finally, we point out that it is possible to create mesoscopic Schr\"odinger's-cat--like
states [NOON-like states \cite{lee}, i.e., states proportional to 
$(|N\rangle +e^{i\theta}|-N\rangle)$, where $\theta$ is a phase] 
using the state-dependent CDT. In this scheme, 
we assume that $U_0N/J_0\ll 1$ and choose $|N\rangle$ as the initial state.
We modulate $U$ at a frequency $\omega$ that corresponds to 
a partial CDT between $|N\rangle$ and $|N-2\rangle$, that is, 
${\cal J}_0\left[(U_1/\omega) (N-1)\right] =0$. At this frequency  
the phases of the Floquet eigenstates $\psi_N^{(\pm)}$,
which are very close to NOON states, become degenerate \cite{note_3fold}.
By detuning from this partial CDT, we have a coherent oscillation 
(with period $T$) between
$\psi_N^{(+)}$ and $\psi_N^{(-)}$. As a result, the initial state $|N\rangle$
evolves into a NOON-like state at $t=T (2n-1)/4$, with $n=1,2,3, ...$.
With increasing the absolute value of the detuning,
the period $T$ decreases but the amplitudes of the components other than
$|\pm N\rangle$ increase, so that the oscillation between the NOON states
is disturbed.  Therefore, $\omega$ (more precisely, $U_1/\omega$) 
should be optimized.
In Fig.\ \ref{fig_noonmodu}, we show the time evolution of 
the normalized population imbalance 
$\langle \Delta N\rangle/N$ and
its variance $\sigma_{\Delta N}/N\equiv N^{-1}\sqrt{\langle \Delta N^2\rangle
-\langle \Delta N\rangle^2}$ for $N=21$ and $N=51$ as examples.
Here 
$\langle \Delta N\rangle \equiv \langle \psi|\Delta \hat{N}|\psi\rangle$ and
$\langle \Delta N^2\rangle\equiv \langle \psi|(\Delta \hat{N})^2|\psi\rangle$
with 
$\Delta\hat{N}\equiv \hat{c}_1^\dagger \hat{c}_1-\hat{c}_2^\dagger \hat{c}_2$. 
These are optimized cases with the amplitude of the wiggles 
in the oscillation of $\langle \Delta N\rangle/N$ being $\alt 0.05$.
When $\langle \Delta N\rangle=0$, $\sigma_{\Delta N}/N$ is almost equal to $1$, which is the largest possible  value; this is a unique property of NOON states.
Note that the oscillation periods are comparable in the two cases:
$T/T_0=211.3$ and $T/T_0=367.4$ for $N=21$ and $N=51$, respectively.
This shows that an advantage of the present scheme is that the optimized $T$
does not increase exponentially with $N$ unlike the tunneling period of 
the higher-order co-tunneling in the self-trapping regime.
This may be understood by the fact that the static part of the
interaction strength $U_0$ is very small ($U_0N/J_0 \ll 1$).
A disadvantage is that we need to  know the number of particles exactly and to fine-tune $U_1/\omega$.
This scheme can be used regardless of the value of $N$ if $U_0 N/J_0\ll 1$.

\section{Creating a NOON state by an adiabatic sweep 
}\label{sec_noon}

\begin{figure}[ht]
\begin{center}
\rotatebox{0}{
\resizebox{8.5cm}{!}
{\includegraphics{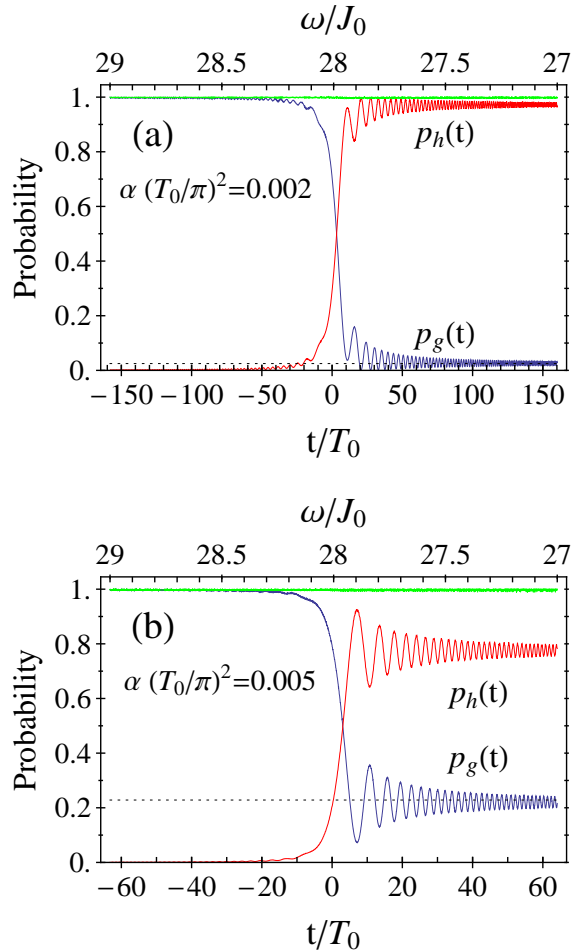}}}
\caption{\label{fig_lz}(Color online) 
Probabilities $p_g(t) \equiv |\langle \psi_g|\Psi(t)\rangle|^2$ [blue (dark gray) lines], 
$p_h(t) \equiv |\langle \psi_h|\Psi(t)\rangle|^2$ [red (medium gray) lines], 
and $p_g + p_h$ [green (light gray) lines]
as a function of time for two different values of the
sweep rate $\alpha$. Here $N=5$, $U_0/J_0=4$, and $A_J=0.5$
(and $V_0=V_1=U_1=0$). 
The dotted lines correspond to the analytical prediction obtained using Eq. (\ref{eq_p0}).
}
\end{center}
\end{figure}

\begin{figure}[ht]
\begin{center}
\rotatebox{0}{
\resizebox{8.6cm}{!}
{\includegraphics{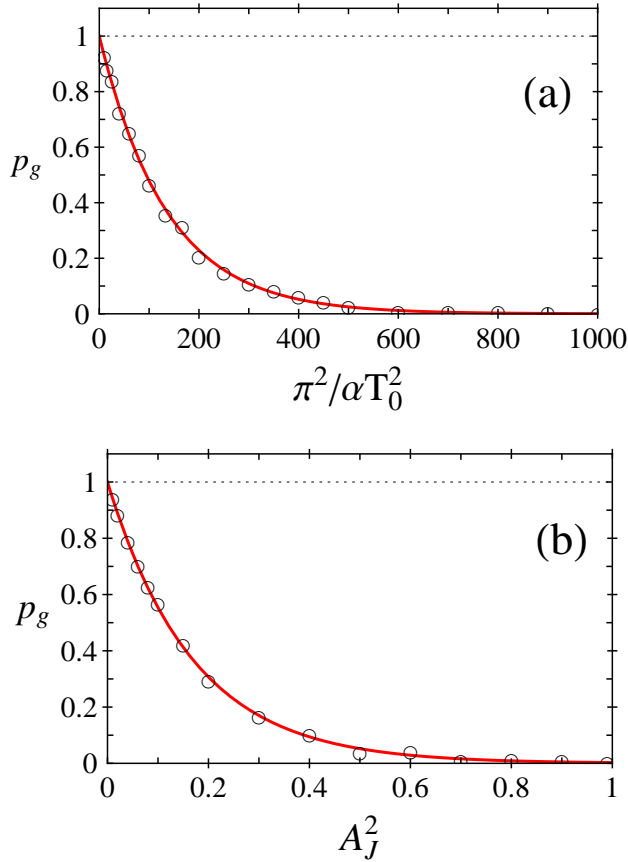}}}
\caption{\label{fig_p0}(Color online) 
Asymptotic value $p_g$ of the transition probability as a function of 
(a) the inverse sweep rate $1/\alpha$ and 
(b) the modulation amplitude $A_J$ for $N=5$ and $U_0/J_0=4$
(and $V_0=V_1=U_1=0$). 
We set $A_J=0.5$ in (a) and $\alpha T_0^2/\pi^2 = 0.005$ in (b). 
The circles show 
the numerical results, and the solid lines show the semi-analytic results
obtained from the Landau-Zener formula (\ref{eq_p0}). The initial time $t_{\rm i}$ 
of the time evolution is chosen such that $\omega(t_{\rm i})/J_0=29$ in Eq. (\ref{eq_sweep}).
}
\end{center}
\end{figure}

In this section we propose another scheme to create NOON-like states. 
This scheme uses an adiabatic sweep of the modulation frequency. 
It enables us to obtain 
NOON-like states with $N\alt 10$ particles starting from the ground
state $\psi_g$ of the time-independent Hamiltonian  $\hat{H}_0$.
The basic idea is to create an avoided crossing between the Floquet
eigenstate corresponding to $\psi_g$ and the one corresponding to 
the NOON-like eigenstate $\psi_h$
by time-periodic modulation,
which changes the geometry of the (quasi)energy space to be periodic.
Here, we modulate the hopping parameter $J$ and set the tilt $V=0$.
Since the phase $\phi_J$ of the modulation does not affect 
the result, we choose $\phi_J=0$ for definiteness. 
The time-independent part $\hat{H}_0$ of the Hamiltonian 
$\hat{H}(t)=\hat{H}_0 + \hat{H}_{T_\omega}(t)$ is given by Eq.\ (\ref{H0}), 
while the time-dependent part $\hat{H}_{T_\omega}(t)$ is
\begin{equation}
  \hat{H}_{T_\omega}(t) = -2J_0 A_J\sin{\omega t}\ \hat{S}_x\ .
\end{equation}
For even $N$, the crossing used in the creation of the NOON state is the one 
between $\psi_N^{(+)}$ and $\psi_{0}^{(+)}$. For odd $N$, it is the one 
between $\psi_N^{(+)}$ and $\psi_{1}^{(+)}$. 
We consider the regime $U_0N/J_0\gg 1$, 
where $\psi_N^{(+)}$ is a NOON-like state.  
The ground state $\psi_g$ of $\hat{H}_0$ corresponds to 
$\psi_{0}^{(+)}$ (even $N$) or $\psi_{1}^{(+)}$ (odd $N$), and
the eigenvalue of $\hat{H}_0$ corresponding to $\psi_g$ is denoted by $E_g$.
Similarly, the NOON-like eigenstate $\psi_h$ of $\hat{H}_0$
corresponds to $\psi_{N}^{(+)}$.  State $\psi_h$ has the 
highest energy among symmetric eigenstates of $\hat{H}_0$, 
and its eigenenergy is denoted by $E_h$.
Because $\hat{H}_0 \sim U_0 N^2 \gg J_0N \sim \hat{H}_{T_\omega}$, 
the eigenstates of $\hat{H}_0$ are almost equal to 
the Floquet eigenstates except near the crossing points.
Therefore, $|\langle \psi_g|\psi_0^{(+)}\rangle|^2\simeq 1$ for even $N$,  
$|\langle \psi_g|\psi_1^{(+)}\rangle|^2\simeq 1$ for odd $N$, and 
$|\langle \psi_h|\psi_N^{(+)}\rangle|^2 \simeq 1$. 
As discussed in Sec.\ \ref{sec_timeindep},
when $\omega$ is decreased from a sufficiently large value,
the first crossing occurs between $\phi_N^{(+)}$ and $\phi_{0}^{(+)}$ for even $N$ and between $\phi_N^{(+)}$ and $\phi_{1}^{(+)}$ for odd $N$ \cite{note_rate}. 
Therefore, in principle, this scheme can be used 
without knowing precisely the total number of particles.
The avoided crossing between the phases $\phi_N^{(+)}$ and $\phi_0^{(+)}$ or $\phi_1^{(+)}$ 
is approximately at $\omega_{\rm res} =E_h-E_g$. In the $N=5$ case discussed earlier, 
this crossing corresponds to the rightmost circle in Fig.\ \ref{fig_floq}(b). 

Let us take $\psi_g$ as the initial state.
If we sweep $\omega$ adiabatically across the avoided crossing, 
$\psi_g$ undergoes an almost perfect transition to $\psi_h$.  
We consider a linear sweep of the form
\begin{equation}
  \omega(t) = \omega_{\rm res} - \alpha t,
\label{eq_sweep}
\end{equation}
where $\omega_{\rm res}$ is the location of the crossing 
and $\alpha$ is the sweep rate. The initial and final times 
of the sweep are denoted by $t_{\rm i}$ and $t_{\rm f}$, respectively. 

In the following calculations, we set $N=5$ and $U_0/J_0=4$. 
The avoided crossing is at $\omega/J_0\simeq 28$.  
In Fig.\ \ref{fig_lz}, we show the time evolution of the probability 
$p_g(t) \equiv |\langle \psi_g|\Psi(t)\rangle|^2$ [blue (dark gray) lines]
at which the system stays in the initial state $\psi_g$ and
the probability $p_h(t) \equiv |\langle \psi_h|\Psi(t)\rangle|^2$ [red (medium gray) lines]
at which the system undergoes a transition to the target state $\psi_h$.
Note that $p_g + p_h$ shown by the green (light gray) lines in Fig.\ \ref{fig_lz} 
is very close to unity throughout the calculations (the deviation is within $0.1$\%), 
and the system is, to a very good approximation, restricted to the subspace 
spanned by the two states.
Thus the crossing can be described by the Landau-Zener (LZ) model \cite{landau,zener,majorana,stuckelberg}.
We denote the modulation period at the crossing point by   
$T_{\rm res}\equiv 2\pi/\omega_{\rm res}$.
The difference between the phases of the Floquet eigenvalues at $\omega(t)$ is
$\Delta\phi =(\phi_{N}^{(+)}-\phi_{0,1}^{(+)})
 \simeq -(E_h - E_g) (T_\omega -T_{\rm res})$. 
Here, we shift the phase difference so that the crossing 
at $\omega\simeq \omega_{\rm res}$ is passed at $t=0$, 
in accordance with the standard expression of the LZ Hamiltonian.
For the linear sweep of Eq.~(\ref{eq_sweep}), we get 
$T_\omega(t) = 2\pi/\omega(t) \simeq (2\pi/\omega_{\rm res}) (1+\alpha t/\omega_{\rm res})$. 
Here we assume that $\alpha t \ll \omega_{\rm res}$. We obtain
the quasienergy separation $\Delta E$ corresponding to $\Delta\phi$ near the crossing as
\begin{equation}
\Delta E= \frac{\Delta \phi}{T_{\rm res}} \simeq - \, \alpha\, t\ ,
\end{equation}
where we have approximated $\omega_{\rm res}\approx E_h-E_g$. 
The diagonal matrix elements $H_h$ and $H_g$ of the LZ Hamiltonian are thus
$H_{h,g} = \pm \Delta E/2$,
where the upper sign corresponds to $H_h$ 
and the lower one corresponds to $H_g$.
We found that the off-diagonal elements $H_{hg}$ and $H_{gh}=H_{hg}^*$ of the effective Hamiltonian are 
to a good approximation given by 
$H_{hg} = - J_0 A_J\langle\psi_h |\hat{S}_x|\psi_g\rangle/\sqrt{2}$.
Consequently, the asymptotic value $p_g$ of the transition probability $p_g(t)$,
$p_g\equiv \lim_{t\rightarrow \infty} p_g(t)$, is \cite{zener}
\begin{align}
p_g =& \exp{\left[ -2\pi \frac{|H_{hg}|^2}{|\partial_t (H_{h}-H_g)|}\right]}\nonumber\\
=& \exp{\left[-\frac{\pi J_0^2 A_J^2|\langle \psi_h|\hat{S}_x|\psi_g\rangle|^2}{\alpha}
\right]}.
\label{eq_p0}
\end{align}

In Fig.\ \ref{fig_p0}, we show the probability $p_g$ as a function of
the inverse sweep rate $1/\alpha$ [Fig.\ \ref{fig_p0}(a)] and
the modulation amplitude $A_J$ [Fig.\ \ref{fig_p0}(b)].
Since $p_g(t)$ and $p_h(t)$ continue to oscillate around the asymptotic value
until far after the crossing (see Fig.\ \ref{fig_lz}),
we calculate $p_g$ by taking the time average of $p_g(t)$ 
after its oscillation amplitude becomes small and almost time independent.
These results are shown by circles in Fig.\ \ref{fig_p0}.
Semianalytical results obtained from Eq.\ (\ref{eq_p0}) are shown
by the red solid lines.  For the parameters used here
($N=5$ and $U_0/J_0=4$), we have $E_g/J_0=12.31$, $E_h/J_0=40.31$, 
$|\langle\psi_h|\hat{S}_x|\psi_g\rangle|=9.697\times 10^{-2}$, and 
$\omega_{\rm res}\approx E_h-E_g =28.00 J_0$.
The agreement between the semianalytical and numerical results is very good.

Finally, we examine the experimental feasibility of this scheme.
According to Eq.\ (\ref{eq_p0}),
to obtain a NOON-like state, we should satisfy the adiabaticity condition:
\begin{equation}
\frac{\pi J_0^2 A_J^2|\langle \psi_h|\hat{S}_x|\psi_g\rangle|^2}{\alpha} \gg 1.
\label{eq_ad}
\end{equation}
In addition, the initial and the final frequencies should be 
outside the crossing region.
Since the range of $\omega$ characterizing the crossing region
is comparable to the level separation $\Delta=2|H_{hg}|$ 
at the avoided crossing,
the initial time $t_{\rm i}$ and the final time $t_{\rm f}$ 
of the sweep have to satisfy
$|\omega(t_{\rm i,\, f})-\omega_{\rm res}|=\alpha |t_{\rm i,\, f}|\agt 2|H_{hg}|$. Also the Landau-Zener formula is valid under this condition.
Taking into account the adiabaticity condition (\ref{eq_ad}), this leads 
to the requirement 
\begin{equation}
|t_{\rm i}|,t_{\rm f} \gg \frac{\sqrt{2}}{\pi^2}
  \frac{T_0}{A_J |\langle \psi_h|\hat{S}_x|\psi_g\rangle|}.
\label{eq_tsweep}
\end{equation}
As an example, let us estimate the time scale given by this equation 
by using the parameters used in the experiment of Ref.\ \cite{foelling}. 
In this experiment, the frequency of the pair tunneling
is $4J_0^2/U_0\simeq 550$ Hz for $U_0/J_0=5$; thus $T_0\simeq 0.72$ ms. 
If $A_J=0.5$, the right-hand side of Eq.\ (\ref{eq_tsweep}) is 
$9$ ms for $N=6$, $40$ ms for $N=7$, and
$214$ ms for $N=8$.
Therefore, a NOON state with $N\alt 7$ could be created
within an experimentally accessible time, 
provided the value of $\omega$ can be controlled 
with a sufficiently high accuracy.
We note that, more generally, an upper limit for $N$
for this scheme to work is $N\simeq 10$.
Since the width of the peaks in the probability distribution (in the Fock space)
of $\psi_g$ and $\psi_h$ scales as $\sim N^{1/2}$,
a few times $N^{1/2}$ should be larger than $N$ in order to 
have an overlap between $\psi_g$ and $\psi_h$ and to have a significant
nonzero value of $|\langle \psi_h|\hat{S}_x|\psi_g\rangle|$.

In the present scheme, the modulation of the hopping parameter works
much better than the modulation of the tilt.  This can be seen using
perturbation theory.  A straightforward calculation shows that for odd
number of particles $\langle\psi_h |\hat{S}_x|\psi_g\rangle \sim
(J_0/U_0)^{(N-3)/2}$ and for even number of particles $\langle\psi_h
|\hat{S}_x|\psi_g\rangle \sim (J_0/U_0)^{(N-2)/2}$.  In the same way,
perturbation theory shows that $\langle\psi_h'
|\hat{S}_z|\psi_g\rangle \sim (J_0/U_0)^{(N-1)/2}$ for odd $N$ and
$\langle\psi_h' |\hat{S}_z|\psi_g\rangle \sim (J_0/U_0)^{N/2}$ for
even $N$.  Here $\psi_h'$ is the antisymmetric eigenstate of
$\hat{H}_0$ with the highest energy.  We see that $|\langle\psi_h'
|\hat{S}_z|\psi_g\rangle|^2/|\langle\psi_h
|\hat{S}_x|\psi_g\rangle|^2\sim (J_0/U_0)^2$, and consequently, the
off-diagonal elements of the LZ Hamiltonian are much smaller when the
tilt is modulated than when the tunneling is modulated.

\section{Conclusions}\label{sec_conclusions}

In this paper, we have considered a time-periodically modulated
two-mode Bose-Hubbard model. We have discussed three types of
modulations, one where the tunneling amplitude is modulated, another where the
interaction strength is modulated, and a third where the energy difference between
the modes (tilt) is modulated.  We focused mainly on the self-trapping
regime, characterized by $U_0N\gg J_0$, and assumed that the amplitude
of the modulation is small.  It is known that a modulation of the
tunneling amplitude can lead to a drastic reduction of the tunneling
period \cite{catform}. We found that a similar effect can be induced
by modulating the interaction strength or the energy difference
between the modes.  We have analyzed this phenomenon using Floquet
theory as the main tool. We found that regardless of the modulated
variable, the system has resonances at some modulation frequencies,
corresponding to either greatly reduced or enhanced tunneling
periods. To a good approximation, the locations of the resonances can
be obtained with the help of the energy eigenvalues of the
time-independent part of the Hamiltonian. Consequently, the locations
of the resonances are almost independent of whether the tunneling,
interaction, or tilt is modulated.

We found numerically that if the tunneling amplitude or interaction strength is
modulated, the system has a wide resonance; that is, the tunneling
period is greatly reduced in a wide range of modulation frequencies.
This resonance is present also in the case of modulated tilt, but it
is much narrower. Furthermore, the behavior of the tunneling period as a 
function of the modulation frequency is not smooth in this case; see Fig.~\ref{fig_modv}. 
These differences can be explained using Floquet theory. The
presence of resonances is related to avoided crossings of the phases
of the Floquet eigenvalues.  In the case of a modulated tunneling matrix
element or interaction strength, the avoided crossings correspond to
Floquet eigenstates with the same parity. In the case of modulated
tilt, these avoided crossings correspond to eigenstates with opposite
parity. In Sec.\ \ref{sec_modv}, it is shown that due to this difference, 
a wide smooth resonance cannot be obtained in the case of modulated tilt.

We have also analyzed cases where the interaction energy is weak in
comparison with the tunneling energy, $U_0N/J_0\alt 1$, and the
modulation amplitude of either the interaction strength or the tilt is
large.  Under these conditions, tunneling can be suppressed at some
specific modulation frequencies. This phenomenon, the coherent
destruction of tunneling, has been extensively studied in the
literature.  We concentrated on a property of CDT that has received
less attention in the previous studies, namely, the Floquet spectrum of
a system where the tilt is modulated.  As expected, we found that the
suppression of tunneling takes place when the phases of the Floquet
eigenvalues become degenerate. For an even number of particles the
suppression is more complete than that for an odd number of particles.

Finally, we have proposed two ways to create a NOON state.  One is
based on coherent oscillation resulting from a detuning from a partial
CDT caused by modulated interaction strength.  An advantage of this
method is that the tunneling period does not increase exponentially
with the total number of particles $N$.  The other method is based on
sweeping the modulation frequency of the tunneling term adiabatically.
This scheme requires neither precise knowledge of the number of
particles nor fine-tuning of the modulation frequency.  We have shown
that by using the latter method and the parameters of a recent
experiment \cite{foelling}, it is possible to obtain NOON states of
$N\alt 7$ particles.

It is known that the mean-field theory of the time-periodically modulated
two-mode Bose-Hubbard model shows chaotic dynamics (e.g., Refs.\ \cite{abdullaev00,holthaus01,lee01,salmond,wang05,mahmud05,weiss,strzys08}).
In the future, it would be interesting to study the connection between
the Floquet spectrum of the original quantum system 
and the chaotic mean-field dynamics.
Another interesting problem to study would be the 
quantum dynamics determined by a 
time-periodically modulated Hamiltonian in the presence of dissipation.
In particular, the engineered dissipation leading to squeezed states 
proposed in Ref.\ \cite{squeeze} is of interest.

\begin{acknowledgments}
We acknowledge Ippei Danshita, Chris Pethick, and 
Sukjin Yoon for helpful discussions.
GW acknowledges the Max Planck Society, the Korea Ministry of
Education, Science and Technology, Gyeongsangbuk-Do, and Pohang City
for the support of the JRG at APCTP. GW is also supported by Basic
Science Research Program through the National Research Foundation of
Korea (NRF) funded by the Ministry of Education, Science and
Technology (No. 2012008028).
\end{acknowledgments}

\appendix

\section{Effective hopping parameter for modulated $J$\label{sect_jeff}}

Here we derive the effective tunneling amplitude  
in the limit of large-amplitude tilt modulation. 
The system follows the Schr\"odinger equation
\begin{align}
i\dot{\psi}(t)=\hat{H}(t)\psi(t),
\label{eq_tdsch}
\end{align}
with
\begin{equation}
\hat{H}(t) = -2J_0\hat{S}_x + U_0 \hat{S}_z^2 + V(t) \hat{S}_z
\end{equation}
and $V(t)$ given by Eq.\ (\ref{eq_vt}).
We go to a rotating system by defining 
\begin{align}
\tilde{\psi}(t)=e^{i\alpha(t)\hat{S}_z}\psi(t), 
\end{align}
where
\begin{align}
\alpha(t) &= \int_{0}^{t}d\tau\,\left[ V_0+V_1 \sin(\omega\tau+\phi_V)\right]\\ 
&= V_0 t+\frac{V_1}{\omega}[\cos\phi_V -\cos(\omega t+\phi_V)].
\end{align}
Using this, the Schr\"odinger equation becomes
\begin{align}
i\dot{\tilde{\psi}}(t)=\tilde{H}(t) \tilde{\psi}(t), 
\end{align}
where
\begin{align}
\tilde{H}(t)&=-2J_0\left(\cos[\alpha(t)]\,\hat{S}_x-\sin[\alpha(t)] \,\hat{S}_y\right)+U_0 \hat{S}_z^2.
\end{align}
Assuming that the modulation period $T_\omega=2\pi/\omega$ is the shortest time scale in the system,
it is possible to obtain an effective Hamiltonian by averaging over $T_\omega$ as
\begin{align}
\tilde{H}_{\textrm{AVE}}(t) &=\frac{1}{T_{\omega}}\int_t^{t+T_{\omega}} \tilde{H}(\tau) \, d\tau\\
&=-2\Jxe(t)\hat{S}_x-2\Jye(t)\,\hat{S}_y + U_0 \hat{S}_z^2. 
\label{eq_heff}
\end{align}
The effective tunneling amplitudes are defined as 
\begin{align}
\Jxe(t)&= \frac{J_0}{T_{\omega}}\int_t^{t+T_{\omega}} \cos[\alpha(\tau)] \, d\tau\\
\Jye(t)&= -\frac{J_0}{T_{\omega}}\int_t^{t+T_{\omega}} \sin[\alpha(\tau)] \, d\tau.
\end{align}
Instead of calculating $\Jxe(t)$ and $\Jye(t)$ separately, we write
\begin{align}
\nonumber 
&\Jxe(t)-i\Jye(t)= \frac{J_0\, e^{i\frac{V_1}{\omega}\cos\phi_V}}{T_{\omega}}\\
&\times\int_t^{t+T_{\omega}} d\tau\, 
e^{i\left[V_0 \tau-\frac{V_1}{\omega}\cos(\omega \tau+\phi_V)\right]}.
\end{align}
This integral can be calculated easily using the equation
\begin{align}
\label{Jnsum}
e^{i z\cos\gamma}=\sum_{n=-\infty}^{\infty}\, {\cal J}_n(z) e^{i n (\gamma+\frac{\pi}{2})},
\end{align}
where ${\cal J}_n(z)$ are Bessel functions of the first kind. We thus obtain 
\begin{align}
&\Jxe(t)-i\Jye(t)\nonumber\\
&=
\begin{cases}\displaystyle
\frac{2 J_0}{T_{\omega}}\sin\left(\frac{\pi V_0}{\omega}\right)e^{i \left[V_0 (t+\frac{\pi}{\omega})+\frac{V_1}{\omega}\cos\phi_V\right]}\vspace{1mm}\\
\displaystyle
\quad\times \sum_{n=-\infty}^{\infty}\, {\cal J}_n\left(\frac{V_1}{\omega}\right)
\frac{e^{in(\omega t+\phi_V-\frac{\pi}{2})}}{V_0+n\omega},&\displaystyle
\frac{V_0}{\omega}\not\in \mathbb{Z}\vspace{3mm}\\
\displaystyle
{J_0 \cal J}_{k}\left(\frac{V_1}{\omega}\right)e^{i\frac{V_1}{\omega}\cos\phi_V} 
e^{-ik (\phi_V+\frac{\pi}{2})},&\displaystyle \frac{V_0}{\omega}=k \in \mathbb{Z}. 
\end{cases}
\label{eq_jeff}
\end{align}
In the special case $V_0/\omega=k\in\mathbb{Z}$, the original tunneling amplitudes $J_x=J_0$ and $J_y=0$ are replaced by effective ones, 
\begin{align}
\Jxe(t) &=J_0 {\cal J}_{k}\left(\frac{V_1}{\omega}\right)\cos\left[k\left(\phi_V+\frac{\pi}{2}\right)-\frac{V_1}{\omega}\cos\phi_V\right],\\ 
\Jye(t) &=J_0 {\cal J}_{k}\left(\frac{V_1}{\omega}\right)\sin\left[k\left(\phi_V+\frac{\pi}{2}\right)-\frac{V_1}{\omega}\cos\phi_V\right],
\end{align}
where $V_1$ is non-zero.

\section{Effective hopping term for modulated $U$\label{sect_umod}}
In the case of large-amplitude modulation of the interaction strength, the
coherent destruction of tunneling is state-dependent \cite{gong09}.
Here, we derive the effective Hamiltonian for this case.

We start from the time-dependent Schr\"odinger equation (\ref{eq_tdsch})
with the Hamiltonian
\begin{equation}
  \hat{H}(t) = -2J_0\hat{S}_x + U(t) \hat{S}_z^2,
\end{equation}
where $U(t)$ is given by Eq.\ (\ref{eq_ut}).  
For simplicity, we set $V=0$.
As in Appendix \ref{sect_jeff}, we go to the rotating frame by defining
\begin{equation}
  \tilde{\psi}(t) = e^{i\alpha(t)\hat{S}_z^2} \psi(t),
\end{equation}
where
\begin{align}
  \alpha(t) =& \int^t_0 d\tau [U_0 + U_1 \sin{(\omega\tau+\phi_U)}]\nonumber\\
  =&\ U_0 t + \frac{U_1}{\omega} \left[\cos\phi_U - \cos{(\omega t+\phi_U)}\right].
\end{align}
Thus the Schr\"odinger equation becomes 
$i\dot{\tilde{\psi}}(t)=\tilde{H}(t)\tilde{\psi}(t)$,
with
\begin{equation}
  \tilde{H}(t) = -J_0\left[ \hat{S}_+ e^{i\alpha(t) (2\hat{S}_z +1)}
+ e^{-i\alpha(t) (2\hat{S}_z+1)} \hat{S}_- \right],
\end{equation}
where $\hat{S}_\pm \equiv \hat{S}_x \pm i\hat{S}_y$.
We have used the equations $[\hat{S}_z^2,\hat{S}_+]=\hat{S}_+ (2\hat{S}_z+1)$,
$[\hat{S}_z^2,\hat{S}_-]=-(2\hat{S}_z+1)\hat{S}_-$, and 
$\hat{S}_x=(\hat{S}_++\hat{S}_-)/2$ to obtain
\begin{align}
  & e^{i\alpha(t)\hat{S}_z^2} \hat{S}_x e^{-i\alpha(t)\hat{S}_z^2}\nonumber\\
  & = \frac{1}{2}\left[ \hat{S}_+ e^{i\alpha(t) (2\hat{S}_z +1)}
+ e^{-i\alpha(t) (2\hat{S}_z+1)} \hat{S}_- \right].
\end{align}
By time averaging over one modulation period $T_\omega$, 
the effective Hamiltonian reads
\begin{align}
  \tilde{H}_{\rm AVE}(t) =& 
\frac{1}{T_\omega} \int^{t+T_\omega}_t \tilde{H}(\tau)d\tau\nonumber\\
=& -J_0 [\hat{S}_+ \hat{A} + \hat{A}^\dagger \hat{S}_-].
\end{align}
Here $\hat{A}$ is defined as
\begin{widetext}
\begin{equation}
  \hat{A}|\Delta N\rangle =
\begin{cases}
\displaystyle
\frac{2}{T_\omega} \sin{\left[\frac{\pi U_0}{\omega}(\Delta N+1)\right]} e^{i\left[U_0(t+\frac{\pi}{\omega}) + \frac{U_1}{\omega}\cos{\phi_U} \right](\Delta N+1)}\vspace{1mm}\\
\displaystyle
\qquad\times\sum_{n=-\infty}^{\infty} 
{\cal J}_{n}\left[\frac{U_1}{\omega}(\Delta N+1)\right]
\frac{e^{-in(\omega t+\phi_U+\frac{\pi}{2})}}{U_0 (\Delta N+1)-n\omega}|\Delta N\rangle,\quad 
&\displaystyle\frac{U_0}{\omega}(\Delta N+1)\not\in \mathbb{Z},\vspace{3mm}
\\
\displaystyle
{\cal J}_{k}\left[\frac{U_1}{\omega}(\Delta N +1)\right]
e^{i\frac{U_1}{\omega} (\Delta N+1) \cos{\phi_U}} e^{-ik(\phi_U+\frac{\pi}{2})}|\Delta N\rangle,\quad 
&\displaystyle\frac{U_0}{\omega}(\Delta N +1)=k \in \mathbb{Z},
\end{cases}\label{eq_a}
\end{equation}
\end{widetext}
where we have used the equation 
$\hat{S}_z|\Delta N\rangle =(\Delta N/2) |\Delta N\rangle$ and  
$\{|\Delta N\rangle\, ;\, \Delta N =-N,-N+2,-N+4,\ldots, N\}$ is the basis of the system. 
In this basis $\tilde{H}_{\rm AVE}$ is a tridiagonal matrix. 
Note that $\hat{A}$, unlike Eq.\ (\ref{eq_jeff}), depends on $\Delta N$. 
If $\langle m-2|\tilde{H}_{\rm AVE}|m\rangle=0$ 
(here we assume $m > 0$ without loss of generality), 
we get $\langle m|\tilde{H}_{\rm AVE}|m-2\rangle=
\langle -m+2|\tilde{H}_{\rm AVE}|-m\rangle=
\langle -m|\tilde{H}_{\rm AVE}|-m+2\rangle=0$. 
In the special case $(U_0/\omega) [(m-2)+1]=k\in \mathbb{Z}$, 
the condition for partial CDT between states $|m\rangle$ and $|m-2\rangle$, $\langle m-2|\tilde{H}_{\rm AVE}|m\rangle=0$, 
can be written as
\begin{equation}
 {\cal J}_k\left[\frac{U_1}{\omega}(m-1)\right]=0.
\end{equation}
If this equation holds, the Hilbert space can be written as a direct sum 
of three uncoupled subspaces, spanned by $\{|N\rangle,|N-2\rangle,\ldots,|m\rangle\}$,  
$\{|m-2\rangle,|m\rangle,\ldots,|-m+2\rangle\}$, and $\{|-m\rangle,|-m-2\rangle,\ldots,|-N\rangle\}$.


\begin{thebibliography}{99}
\bibitem{note_crossings} As we shall see later, for odd $N$ these $N-2$ crossing types
consist of one type where $\phi_{|\Delta N|=N-2}^{(\pm)}$ crosses $\phi_{|\Delta N|=N}^{(\pm)}$ 
successively and $N-3$ types where $\phi_{|\Delta N|=N-4}^{(\pm)}\ , \cdots ,\
\phi_{|\Delta N|=1}^{(\pm)}$ crosses $\phi_{|\Delta N|=N}^{(\pm)}$ once.
For even $N$, there is one crossing type with $\phi_{|\Delta N|=N-2}^{(\pm)}$
and $N-4$ types with each $\phi_{|\Delta N|=N-4}^{(\pm)}\ , \cdots ,\
\phi_{|\Delta N|=2}^{(\pm)}$ and one type with $\phi_{\Delta N=0}^{(+)}$.
In the case of modulated $J$ ($V$), the phase of the Floquet eigenvalue 
$\phi_{\Delta N=0}^{(+)}$ undergoes avoided crossings with the phases of the eigenstates of even (odd) parity.
\bibitem{note_narrowres} Based on the behavior of the phases of the Floquet eigenvalues,
there are resonances also at $\omega/J_0 \simeq 11$ and $14$. These 
are too narrow to be observed in Fig.\ \ref{fig_modv}.
\bibitem{note_u0=0} For $U_0=0$, the effective Hamiltonian 
[see Eq.\ (\ref{eq_heff})] vanishes when the condition for
CDT is satisfied: ${\cal J}_0(V_1/\omega)=0$. When this equation holds,
the phases of all the Floquet eigenvalues vanish, and the degeneracy is complete irrespective of
whether $N$ is even or odd.
\bibitem{note_3fold} For even $N$ this degeneracy involves
three states if $U_0=0$: $\psi_N^{(\pm)}$ and another Floquet eigenstate
whose quasienergy is identically zero. The third state disturbs
the desired oscillation between $\psi_N^{(\pm)}$.
The third state can be easily lifted to change the threefold degeneracy 
to a twofold one by introducing nonzero, but small, $U_0$ such that
$U_0N/J_0 \ll 1$.
\bibitem{note_rate} This can be understood by the fact that, 
with decreasing $\omega$, 
$\phi_{0}^{(+)}$ for even $N$ and $\phi_{1}^{(+)}$ for odd $N$
deviate from zero at the lowest rate and $\phi_N^{(+)}$ deviates
at the highest rate among symmetric states.
\end{thebibliography}
\end{document}